\documentclass[traditabstract]{aa}  
\usepackage{graphicx}
\usepackage{txfonts}
\usepackage{epstopdf}
\newfont{\myfont}{cmmib10}

\newcommand{\btheta}{\hbox{\myfont \symbol{18} }}

\newfont{\myfontsmall}{cmmib8}

\DeclareSymbolFont{cmmi}{OML}{cmm}{m}{it}
\DeclareMathSymbol{v}{\mathalpha}{cmmi}{"76}

\begin{document}
   \title{The intra-hour variable quasar J1819$+$3845: 13-year evolution, jet polarization structure and interstellar scattering screen properties
   }

   \author{A.G. de Bruyn\inst{1,2}
          \and
          J.-P. Macquart\inst{3,4}
          }

   \offprints{A.G. de Bruyn.  The data used to construct most of the lightcurves and images in this paper are available via anonymous ftp to CDS at cdsarc.u-strasbg.fr (130.79.128.5).}

   \institute{ASTRON, Postbus 2, 7990 AA Dwingeloo, The Netherlands \email{ger@astron.nl}
         \and
         Kapteyn Astronomical Institute, University of Groningen, P.O. Box 800, Groningen 9700 AV, The Netherlands
         \and
   ICRAR/Curtin University of Technology, Bentley, WA 6845, Australia
              \email{J.Macquart@curtin.edu.au}
          \and
         ARC Centre of Excellence for All-Sky Astrophysics (CAASTRO)
             }

   \date{}

\titlerunning{The long-term evolution of J1819$+$3845}
\authorrunning{de Bruyn \& Macquart}

 
  \abstract{We examine the long-term evolution of the intra-hour variable quasar, J1819$+$3845, whose variations have been attributed to interstellar scintillation by extremely local turbulent plasma, located only 1--3\,pc from Earth.   The variations in this source ceased some time in the period between June 2006 and February 2007.  The evolution of the source spectrum and the long-term lightcurve, and the persistent compactness of the source VLBI structure indicate that the cessation of rapid variability was associated with the passage of the scattering material out of the line of sight to the quasar.  We present an extensive analysis of the linear polarization variations and their relation to total intensity variations.    The proper motion of polarized features in the quasar jet is found to be subluminal.  Systematic time delays between Stokes $I$, $Q$ and $U$, in combination with the structure of the source obtained from 8.4\,GHz VLBI data confirm the estimate of the screen distance: 1--2\,pc, making the screen one of the nearest objects to the Solar System. 

We determine the physical properties of this scattering material.  The electron density in the scattering region is extremely high with respect to the warm ionized ISM, with an estimated density of 
$n_e \sim 97 \, l_0^{1/3} {\Delta L}_{100}^{-1/2}$cm$^{-3}$, where $l_0$ is the outer scale of the turbulence in AU and $\Delta L = 100\, \Delta L_{100}\,$AU is the depth of the scattering region.  If this plasma is in pressure balance with the local magnetic field, one expects a $\sim 2\,$rad\,m$^{-2}$ rotation measure change associated with the passage of this material past the quasar.  To that end, we examine the rotation measures of sources and the diffuse polarized emission in the surrounding region.   We place a limit of 10\,rad\,m$^{-2}$ on the RM change based upon 21\,cm polarization observations.  The variability of sources near J1819$+$3845 is examined to deduce limits on the transverse extent of the screen; we find that no other sources exhibit variations on comparable timescales and that the screen must therefore be either very small (of order 100\,AU) or patchy. 

}  

\maketitle

\section{Introduction}

Over 56\% of all compact flat-spectrum radio sources exhibit Intra-Day Variability (IDV) at centimetre wavelengths (Lovell et al.\,2008).  It is now well-established that the dominant cause of this variability is interstellar scintillation, caused when the turbulent plasma in the Inter-Stellar Medium (ISM) randomly amplifies the radiation from sufficiently compact ($\lesssim 50\,\mu$as) sources (e.g. Dennett-Thorpe \& de Bruyn 2002; Lovell et al.\,2008).  The IDV phenomenon appears to be highly intermittent; the four-epoch MASIV survey of IDV sources found that only 27-36\% of the total sample (i.e. roughly half of the variable sources) exhibited IDV during any one epoch.  Such intermittency is also observed in some well-studied individual sources, wherein bursts of scintillation lasting only a few months and ceasing on a timescale of a few weeks are observed (Kedziora-Chudczer 2006).

The origin of this intermittency is a fundamental and open question.  It may arise either as a result of  evolution in the internal compact structure of the source itself, or changes in the properties of the turbulence along the line of sight to the source.  

AGN are known to exhibit outbursts, associated with the presence of compact structure, on timescales of months to years at radio wavelengths, and the emergence and expansion of such structure could relate to the episodic nature of IDV.   However, this explanation encounters difficulties if it is to explain the disappearance of IDV over timescales of a month or less because of the extreme source properties required.  For a source at $z \sim 1$ to expand sufficiently that its size exceeds the angular scales on which interstellar scintillation occurs at centimetre wavelengths, $\sim200\,\mu$as, one must appeal to apparent source expansion speeds exceeding $\sim40\,c$, which is implausible (Liu \& Zhang 2007; Lister 2003; Vermeulen \& Cohen 1994).

The alternate explanation is that the disappearance of IDV is related to the finite spatial extent of the turbulent patches responsible for the scintillation, so that the variability lasts only as long as it takes for the turbulent plasma ``cloud'' to cross the line of sight. Another related possibility is that the turbulence that gives rise to the scintillations (i.e. turbulence on transverse length scales $\sim 10^8$\,m) either dissipates or evolves substantially on a timescale comparable to the disappearance of IDV.  Extrinsic explanations such as these are partially supported by the fact that some IDV sources (e.g. PKS\,0405$-$385; Kedziora-Chudczer 2006) exhibit bursts of intra-day variability that appear to bear no relation to the underlying intrinsic behaviour of the source.  However, this explanation then implies that the interstellar medium is pervaded by tiny-scale, $\sim10\,$AU, patches of intense turbulence 
whose presence is at variance with a model based on Kolmogorov-like turbulence that is distributed homogeneously throughout the ISM (Armstrong, Rickett \& Spangler 1995).  There is speculation that such patches may be related to the ionized clouds implicated in Extreme Scattering Events (ESEs; Fiedler et al.\,1987; Walker 2007). Indeed, the accumulated evidence from ESEs, IDV intermittency, pulsar fringing events and from pulsar speckle imaging (e.g. Wolszczan \& Cordes 1987; Brisken et al. 2010) all point to the existence of tiny-scale structure in the ISM.  Studies have shown that a fraction of this material must be overpressured with respect to the ambient warm ionized ISM (e.g. Walker \& Wardle 1998; Jenkins \& Tripp 2001).  The issue now relates to the properties of these structures, namely their distribution, the means by which they are generated, and their internal structure.

The quasar J1819$+$3845 is the most intensely studied of all the intra-day variable sources (Dennett-Thorpe \& de Bruyn 2000, 2002, 2003; Macquart \& de Bruyn 2006, 2007). 
 Its variability, exceeding an rms of 30\% of the mean flux on timescales less than 20\,minutes at a wavelength of $6\,$cm, makes it the most extreme variable extragalactic radio source known, and the prototype of a small group of sources called intra-hour variable sources.   Frequent monitoring observations have established that the source undergoes an annual cycle in the timescale of variability, caused by Earth's motion about the Sun relative to the turbulent ISM (Dennett-Thorpe \& de Bruyn 2003).  This, combined with measurements in the delay of the pattern arrival time between WSRT and the VLA, constrain both the peculiar velocity of the scattering screen responsible for the scintillations, along with the anisotropy and associated position angle of the scintillation pattern.  Power spectral analysis of the $6\,$cm lightcurves have revealed information on the micro-arcsecond structure of the source and shown that it is temporally evolving on a timescale of about a year (Macquart \& de Bruyn 2007).  Most importantly, it has also enabled a measurement of the location of the scattering screen, showing that the extremely rapid variations arise in a region only $1-3$\,pc from Earth (Dennett-Thorpe \& de Bruyn 2003; Macquart \& de Bruyn 2007).   

Sometime in the interval between June 2006 and February 2007,
the rapid scintillations ceased (Cim\`o 2008).  This was one year after the termination of 7.5 years of high-cadence monitoring.  Since then, infrequent monitoring  has only detected very small, $1$--$2\%$ rms, modulations (see also Koay et al.\,2011) on timescales greater than hours to days, and there has been no recurrence of the large, rapid variations that were observed in the source between 1999 and 2006.

The quality and extent of the data on J1819$+$3845 puts us in the uniquely strong position of being able to isolate the cause of this termination.  Were the cause intrinsic to the source, it would enable us to constrain the lifetime of compact IDV components, whereas if the cause were extrinsic it would enable us to confidently deduce a number of physical properties of the scattering screen responsible for the IDV.

In this paper we determine whether the discontinuity in fast variations is attributable to an intrinsic or extrinsic cause.  In \S\ref{sec:obs} we  briefly summarize previous WSRT observations and present the extensive body of old and new data on J1819$+$3845.  In \S\ref{sec:SrcVsScr} we examine the long-term source behaviour, in both total intensity and polarization, and assess the evidence in support of the two possible explanations. We also present a rough picture of the evolving sub-mas jet structure of the quasar.   In \S\ref{sec:SrcProp}  we  discuss observations of the flux density stability  of faint compact background sources surrounding J1819$+$3845 and  analyse the diffuse 21\,cm polarized emission from our Galaxy seen towards J1819$+$3845.  We deduce new limits on the properties of the scattering screen and discuss its implications for the presence of tiny-scale ionized turbulent structures in the ISM.  Our conclusions are presented in \S\ref{sec:con}.

%

\section{Observations} \label{sec:obs}

Most of the data reduction procedures employed for the data presented have been described extensively in previous papers (Dennett-Thorpe \& de Bruyn 2003; Macquart \& de Bruyn 2006; 2007).  We present here a brief summary.  All the observations presented in this paper were obtained with the multi frequency 
frontends located at the prime focus of the Westerbork Synthesis Radio telescope. The frontends cover broad frequency ranges around 350\,MHz (henceforth 85\,cm), 1400\,MHz (21\,cm), 2300\,MHz (13\,cm), 4900\,MHz (6\,cm) and 8500\,MHz (3.6\,cm).


The data presented in this paper, and previous papers, now span a period exceeding 13 years.
In this period the correlator backend changed as described in Dennett-Thorpe \& de Bruyn (2003).  Before April 2002, all data were taken with a total bandwidth of 80\,MHz, covered in eight bands of 10\,MHz.  
After this date, the receiving bandwidth was 80\,MHz at 85\,cm and 160\,MHz at all other frequencies. Some 21\,cm observations were even conducted in frequency mosaicing modes with three to four adjacent 160\,MHz settings covering the band from 1200-1800\,MHz (see Macquart and de Bruyn 2006).  The total 80/160\,MHz band is covered in eight sub-bands of either 10 or 20\,MHz.  
In the observations conducted since April 2002 each subband again is covered with 
64--128 spectral channels. All observations were recorded with full polarization information.  
The integration times varied between 10 and 60 seconds.



The calibration of the data, especially those at 6\,cm and 21\,cm, has been described in previous papers (Dennett-Thorpe \& de Bruyn 2003;  Macquart \& de Bruyn 2006). The large variations shown by J1819$+$3845, however, make our  conclusions insensitive to the precise accuracy of the flux density monitoring.  As detailed in the two papers cited above, the typical flux density accuracy achieved on hourly and daily timescales is about 1--2\%. Although the flux density monitoring of J1819$+$3845 was conducted at a range of frequencies, only at 6\,cm do we have a continuous high-cadence record, spanning a period of 7.5 years. The 6\,cm observations were conducted at typically monthly intervals. Monitoring at all other WSRT frequencies, however, was considerably more irregular.  Some details are given below, where necessary.

%
%

\subsection {3.6 and 13\,cm observations}
\begin{figure}[h]  
 \centerline{\includegraphics[width=85mm]{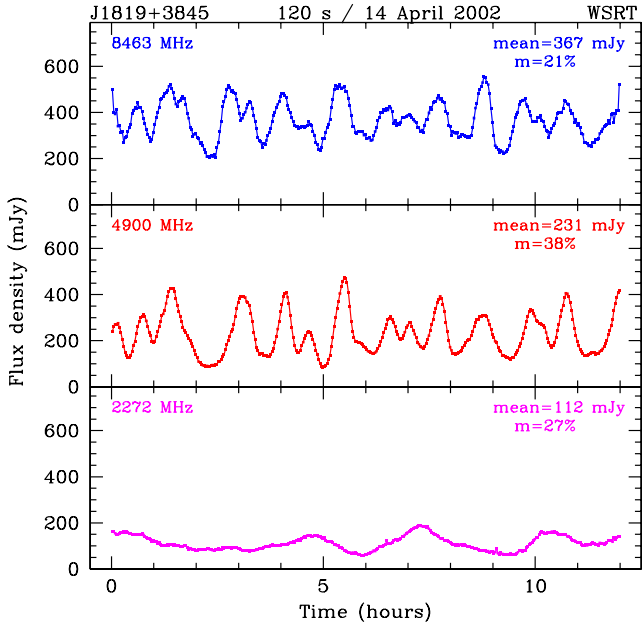}}
 \caption{Variations displayed by J1819$+$3845 at 13, 6 and 3.6\,cm wavelengths on 14 April 2002. The averaging time was 120 seconds. Note the good correlation between the lightcurves at 3.6 and 6\,cm, where the scintillations are in the weak regime.} \label{fig:typicalVar} 
 \end{figure}

All previous papers on this source (Dennett-Thorpe \& de Bruyn 2000, 2002, 2003; Macquart \& de Bruyn 2006, 2007) have concentrated on results at 6\,cm and 21\,cm.  We refer to those papers for readers interested in the details of calibration, processing and presentations of the data.   
However, at many epochs in the period 2001--2006, we conducted simultaneous multi-frequency observations, by splitting the WSRT in sub-arrays of 5 (3.6\,cm), 4 (6\,cm) and 5 (13\,cm) telescopes.  Figure \ref{fig:typicalVar} presents a set of variations exhibited by J1819$+$3845 at 13, 6 and 3.6\,cm during its period of rapid variability in the second quarter of 2002. It shows the typically good correlation between the variations at 3.6 and 6\,cm, while at frequencies lower than 3\,GHz the variations are uncorrelated.   Some specific results on the differential time delay between the scintillations at 3.6 and 6\,cm are presented in the discussion of the VLBI structure in \S\ref{sec:SrcProp}. 

\subsection {21\,cm observations}

\begin{figure}[h]  
 \centerline{\includegraphics[width=85mm]{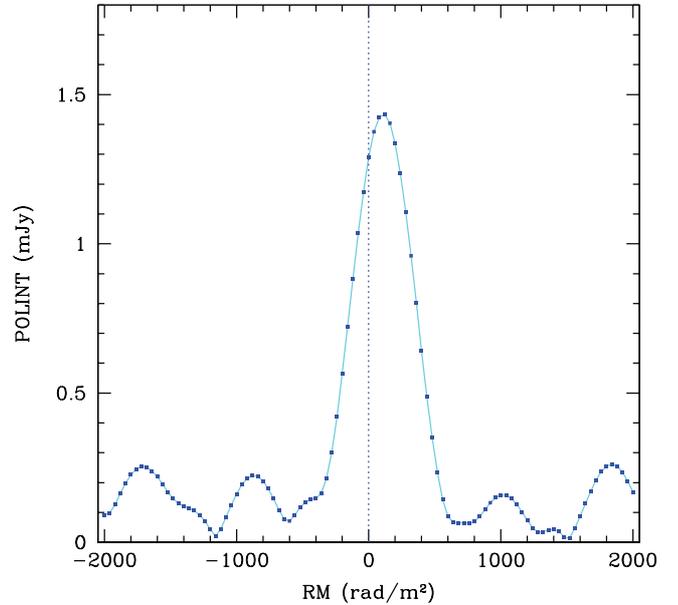}}
 \caption{The Faraday spectrum formed from the polarized emission of J1819$+$3845 in the 21\,cm band from observations on 15 August 2004. The spectrum is sampled with a step of 40 rad m$^{-2}$.} 
 \label{fig:FaradaySpectrum} 
 \end{figure}

A large number of 21\,cm observations were presented in Macquart \& de Bruyn (2006). 
Monitoring of the source, however, started much earlier than the 2002 lightcurves presented in that paper and was continued, less frequently, up until our most recent epoch in June 2012.  We
used these data to search for linear polarization from
J1819$+$3845 and other discrete sources in its immediate neighbourhood
as well as from the diffuse Galactic foreground emission.  For all
21\,cm polarization observations we used standard polarization
calibration procedures. Subsequently we used Rotation Measure (RM) synthesis (Brentjens \& de Bruyn 2005) to produce RM cubes over a range in Faraday depth from typically -500 to +500 rad\,m$^{-2}$.   We detected clear polarization from J1819$+$3845 in about half of all epochs,
especially when the source was bright (due to scintillation). An
example of a Faraday spectrum of J1819$+$3845 is shown in 
Fig.~\ref{fig:FaradaySpectrum}. The peak intensity is located at an RM of
115 rad\,m$^{-2}$. The resolution in Faraday space, given by the width of the RM Spread Function (RMSF), is about 350 rad\,m$^{-2}$. The observed width, however,  is significantly larger,
which may be due to calibration errors. It is also possible, and perhaps even likely, 
that different components in the source have a different RM, which would lead to a broadening of the Faraday spectrum.   

A subset of four 12-hour syntheses, selected for good short-spacing
coverage, was combined to search for diffuse large-scale linear
polarization from our Galaxy.  These data  are presented and discussed in \S\ref{sec:SrcProp}.  

\subsection{85\,cm observations} 

At 85\,cm we conducted two 12\,h observations, on 10 July 2004 and 23 April 2005. 
In both observations the source was detected as a point source with flux densities of
59$\pm$2 and 27$\pm$2 mJy, respectively.   We did not search for
intra-hour variability.   On the basis of scaling laws of refractive
scintillation for both the modulation index and the timescale (Narayan 1992) we do not expect to be able to detect any intra-hour variations.  J1819$+$3845 was also detected, at 22$\pm$4\,mJy, in the WENSS survey (Rengelink et al., 1997). The WENSS flux density, which is referenced to a wavelength of 92\,cm (325\,MHz), represents an average of six 12\,h syntheses in the period from October 1995 to March 1996. (On-line access to the catalogue and images from WENSS are available at http://www.astron.nl/wow.) The source spectrum remains inverted down to very low
frequencies.  The source spectrum at two well-covered epochs is discussed further
in \S\ref{sec:SrcVsScr}.

\section{The cessation of variability: source evolution or finite scattering screen?} \label{sec:SrcVsScr}

In this section we examine several aspects of the variability of J1819$+$3845 that shed light on the cause of the termination of scintillations in the source.  In particular, we examine long-term trends in the variability of the source, and its relation to the compactness of the emission as probed by VLBI.  We also describe the polarization of the source and show that its variability properties evolve over a timescale of months to years.  This is used to place limits on the apparent speed of components along the jet.

\subsection{Long-term behaviour: 1999--2006 and 2007--2012}

Figure \ref{fig:FluxTrend} shows the intrinsic evolution in the flux
density and modulation index source based on 13 years of observations at
6\,cm.  A long-term trend is evident in both the flux density
and modulation index.  The source exhibited a decline in the
modulation index from an average value of 0.35 in the period
1999--2003 to a value of 0.25 in the period 2002--2005,
before rising again to a value of 0.30 in 2006.  However, the drop 
in modulation index that  occurred sometime between June 2006 and 
February 2007 was remarkable. We return to this shortly, after 
addressing the properties of the source in the 7.5 years of rapid
scintillation prior to this date.    

It is clear that the variability properties of the source
undergo long-term evolution. This must at least partially reflect
underlying structural changes in the source, as would be expected in such an extremely compact object.  Scintillation-deduced measurements of the angular size of components in J1819$+$3845 are 
$20-30\,\mu$as (Macquart \& de Bruyn 2007 based on the derived screen
distance of $2.0\pm0.3$\,pc), which corresponds to a transverse scale
of 0.12--0.18\,pc at the source redshift of $0.54$ (we assume a
$\Lambda$CDM cosmology with $\Omega_\Lambda$=0.73, $\Omega_m=$0.27 and H$_0$=71 km\,s$^{-1}$\,Mpc$^{-1}$).

\begin{figure}[h]  
 \centerline{\includegraphics[width=80mm]{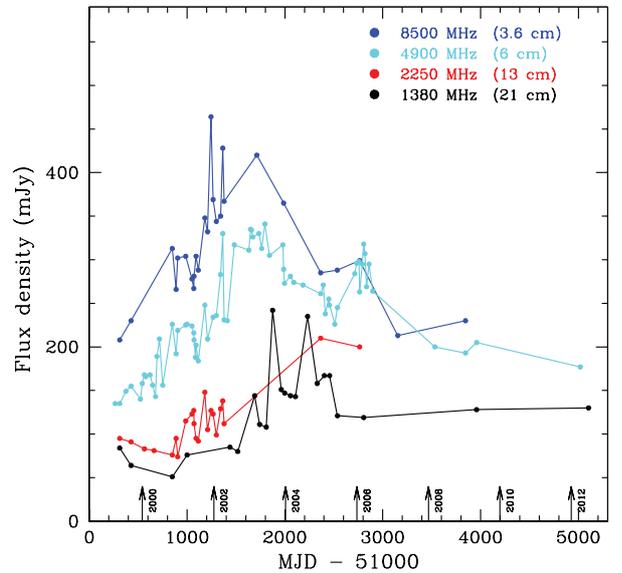}}
 \caption{The lightcurves of J1819$+$3845 at wavelengths of 3.6, 6, 13 and 21\,cm using all available data between 1999 and 2012.5. All flux density data were obtained with the WSRT with exception of the following three epochs: February 2007-8, 8.4\,GHz-VLBA (Cim\`o 2012) and January 2009, 5-8\,GHz-VLA (Koay et al.\,2011).}
 \label{fig:4freqEvolution} 
 \end{figure}

\begin{figure}[h]  
 \centerline{\includegraphics[width=80mm]{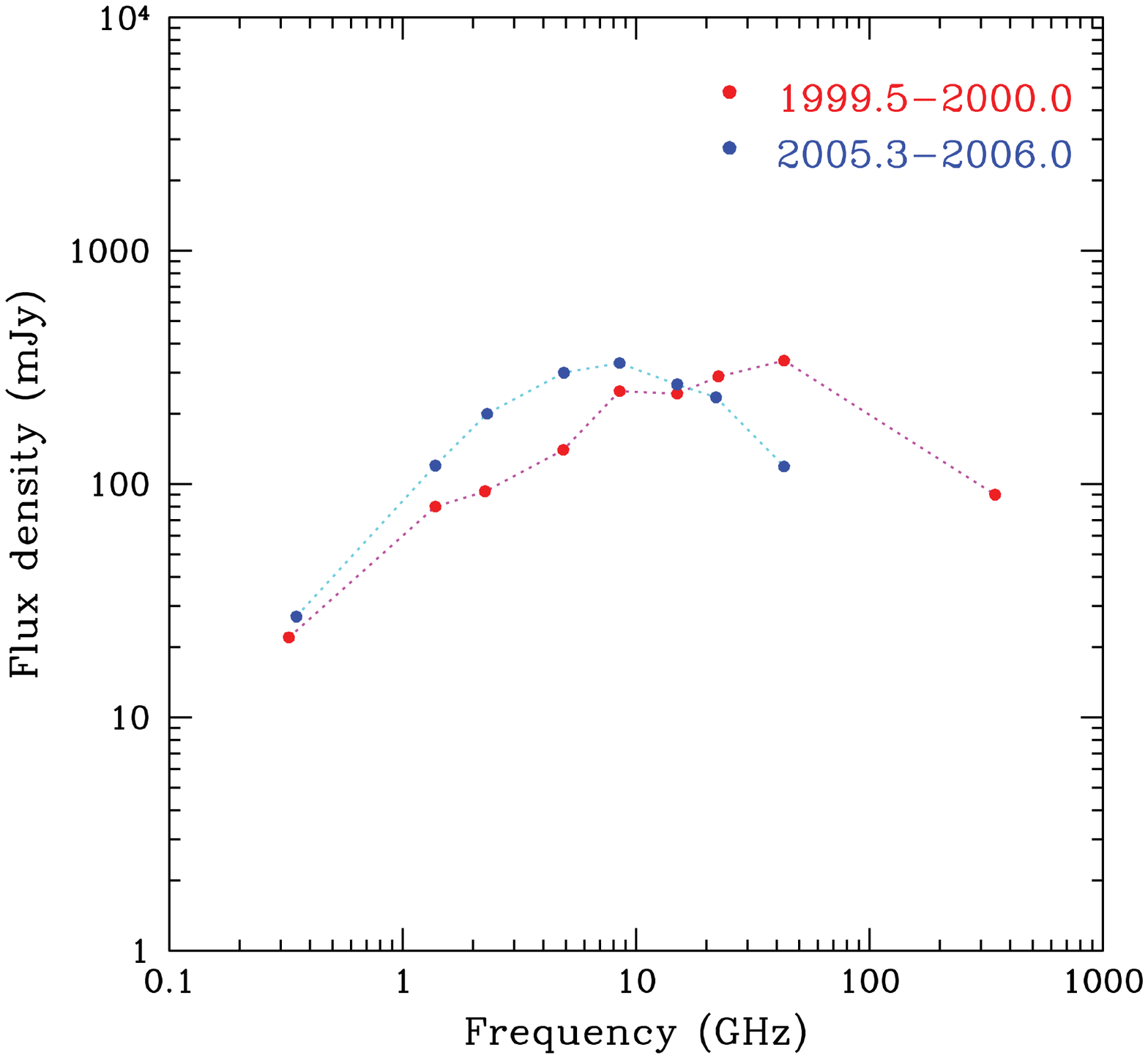}}
 \caption{The broadband spectra of J1819$+$3845 for two
   epochs around 2000.0 and 2006.0. At most frequencies data at
   closely contemporaneous dates could be used. Data used in the
   construction of these spectra came from WSRT, VLA, VLBA and JCMT
   observations presented in this paper, and from Moloney (2010, his Table 3.3, 8--43 GHz, VLBA).
   The lowest frequency data were taken in 1996 (92\,cm) and 2005.3
   (85\,cm).The uncertainties on the data points are dominated by flux scale errors, which are typically a few per cent.} 
 \label{fig:1999p2006Spectrum} 
 \end{figure}

There is some evidence to suggest that occasional intrinsic variations may be present on
timescales of weeks in 2002.  Observations on 1 Apr 2002 
(see de Bruyn, 2012) recorded a flux density of 330\,mJy
at  6\,cm with a modulation index of 43\%. The three highest flux
densities ever recorded in fact took place during that observation.   
A subsequent observation on 14 Apr 2002, which is shown in Fig.\,\ref{fig:typicalVar}, measured a mean flux density of only 231\,mJy and a similar modulation index of 38\%. Because both epochs contained 
a large number ($>12$) of scintillation peaks, the estimated error in each mean flux density is below 10\%, so the difference between the average flux 
density on the two epochs, 330\,mJy versus 231\,mJy, is significant.  An obvious interpretation of this difference is that occasional rapid intrinsic variability occurs in the quasar.  However, it is conceivable that such changes might also represent some form of intermittent longer term scintillation behaviour; variations in the amplitude and duration observed are consistent with ESEs observed in some IDV sources (e.g. Senkbeil et al. 2008).
In general, however, the long-term changes in this 
ultra-compact source are surprisingly slow. A slow but steady 
five-year rise in the flux density, from its discovery in January 1999 to the end of 2003, 
was followed  by a slow but steady decline until the middle of 
2005. This was again followed by a small rise before dense monitoring was terminated. Since
February 2007 the source has continued to fade slowly until the last observations in
June 2012.  A graph showing four lightcurves constructed from the
1.4--8.5\,GHz coverage with the WSRT in the years between 1999 
and 2006 is shown in Fig.~\ref{fig:4freqEvolution}. 
During the first 7.5 years the spectrum remained inverted ($\alpha =+0.8$) between frequencies of 1.4 and 8.5\,GHz, although it gradually became less steep. In recent years the spectrum has almost become flat at 5\,GHz.  Extremely broadband spectra of J1819$+$3845 are shown in Fig.\,\ref{fig:1999p2006Spectrum} for two epochs around 2000 and 2006. 

We have also searched archival data for evidence of past scintillation activity.  J1819$+$3845 was observed twice with the Greenbank telescope at 6\,cm in November 1986 and October 1987 (Gregory et al.\,2001). The average flux density in those years was 91$\pm$9\,mJy, with some evidence for significant variation between these two epochs.  Such a variation would be consistent with a source exhibiting scintillations in the period between these two measurements.  We can not, therefore, determine with certainty when the source commenced scintillating. 

\subsubsection{Cessation of variations}

We now consider the source behaviour after June 2006. 
There was clearly no indication in the 6\,cm mean flux density, nor in the overall spectrum or compact source structure (Moloney 2010), that forewarned of the termination of variability in the source.
The extreme decrement in the modulation index observed on
23 Feb 2007 is unprecedented in the behaviour of this and other
sources, although significant variations in the
modulation index have also been observed in PKS~0405$-$385 (Kedziora-Chudczer 2006). 
The scintillations were first noted to have ceased during the analysis
of the EVN+WSRT 5\,GHz observations in March 2008 (Cim\`o 2008).  
Subsequently, a 8\,GHz VLBA observation of this source taken one year 
earlier, confirmed the result (see Cim\`o 2012). 

\begin{figure}[h]  
\centerline{\includegraphics[width=85mm]{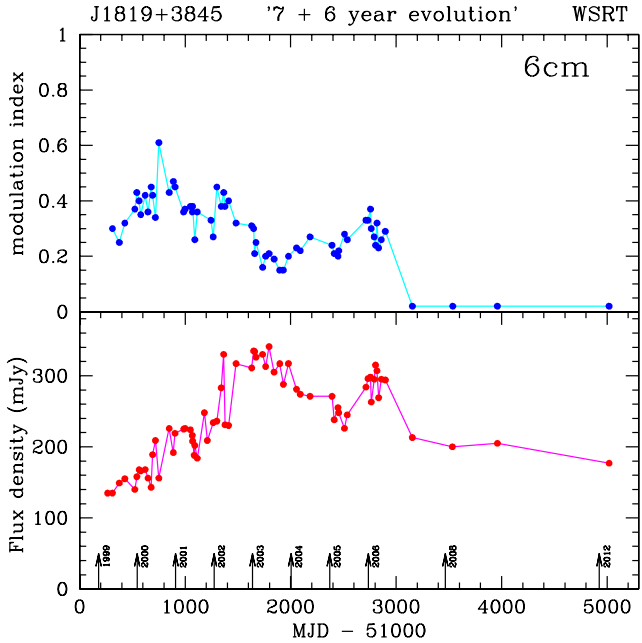}}
\caption{The long term trend in the mean flux density and modulation
index at 4.9\,GHz.  Determining the intrinsic flux density and
modulation index of a scintillating source requires a measurement that
extends over a large number of independent realisations of the
scintillation process. Monte Carlo simulations (Dennett-Thorpe and de
Bruyn 2003) suggest that the relative error in the flux density
decreases as $\approx 1/3N^{-1/2}$ where $N$ is the number of peak-to-peak 
variations, or scintles, exhibited by the lightcurve. To
minimize errors in the long-term lightcurve we therefore only include
data taken in observations extending over at least 6 scintles. The
flux density and modulation index observed in the 8\,GHz VLBA
observation on 23 Feb 2007, were scaled to 4.9\,GHz using the 
average WSRT values observed at those frequencies in the preceding 
3 years. The 2009.0 data points are from Koay et al.\,(2011).
} \label{fig:FluxTrend} 
\end{figure}

Both the 2007 and 2008 datasets place an upper
limit on the source variability of 1\%. 
WSRT observations since then, and VLA results by Koay et al. (2011), have placed comparable upper limits of 1\% on the hourly modulation index at 4.9\,GHz over the period 2009--2012.  This behaviour is in contrast with variations with an average modulation index of 0.3 throughout 2006 prior to the abrupt termination of the variability (see Fig.~\ref{fig:FluxTrend}). For a montage  
of the variations observed in the final seven months, the `Swan Song' of J1819$+$3845, we refer the reader to de Bruyn (2012).   

If the drop in modulation index were attributed to structural changes
in the scintillating source, it would require a large and abrupt
change in the apparent source size.  In the regime of weak
scintillation, applicable to the variations at 6\,cm, and assuming
a Kolmogorov spectrum of turbulence\footnote{Although the existence of compact clouds, such as those responsible for ESEs and the scintillations of IDVs like J1819$+$3845, are at variance with the model of Kolomogorov turbulence, the internal turbulence of these clouds is often well modelled using a power-law spectrum of turbulent density fluctuations whose index is close to the Kolmogorov value.  Scintillation power spectrum modelling shows this to be the case for both J1819$+$3845 (Macquart \& de Bruyn 2007) and other well-studied IDV sources (Rickett, Kedziora-Chudczer \& Jauncey 2002).}, the modulation index for a source
of angular extent $\theta$ scales as $m \propto \theta^{7/6}$.  Thus,
a reduction in the modulation index from $29\%$ to $<$1\% 
would require an increase in source size
by a factor exceeding 18.  This places a conservative lower limit on
the present source size of $360\,\mu$as.  Given that the
scintillations ceased within a period of less than 257 days, this
would require an apparent expansion of the {\it entire} source with a
speed of at least 10\,$c$.  Although individual component motions with
such apparent speeds are observed in AGN, the expansion of an entire
source at this speed would be completely unprecedented.
Moreover, the March 2008 8.4\,GHz VLBA observations in which the source scintillations were noted to have
ceased indicate that the source is unresolved on the longest
observed baseline (about 5700 km).  The visibility
amplitude on this baseline was $210 \pm 15\,$mJy, the same value that
was measured on shorter baselines (Cim\`o, 2012).  We can use this to place a limit on the source size
subsequent to the cessation of variability.  For a source with a
two-dimensional brightness distribution $I(\btheta) = I_0
\exp(-\btheta^2/2 \theta_0^2)/2 \pi \theta_0^2$, where $2 \theta_0
\sqrt{2 \log 2}$ is the full-width at half-maximum (FWHM) of the
source, the corresponding visibility amplitude on a baseline ${\bf r}$
is
\begin{eqnarray} V({\bf r}) = I_0 \exp \left( - \frac{1}{2} k^2 r^2
\theta_0^2 \right).
\end{eqnarray} At the 3-$\sigma$ level we can thus constrain the
source FHWM size to be no larger than $350\,\mu$as.  This constraint
makes it highly implausible to attribute the large drop in modulation
index to an expansion in source size.

        
\subsection{Evolution of polarized components in the jet} \label{subsec:polnComponents}
Polarized emission from J1819$+$3845 at a wavelength of $6\,$cm was
observed as early as 2000 (see de Bruyn \& Dennett-Thorpe 2001) but
had too low a S/N to interpret properly.  Between 2000 and 2002 the 
total flux density increased significantly and the polarized flux 
density along with it.  Typical peak flux densities in Stokes $Q$ and $U$ since 2002 are 4--5 mJy, i.e. about 1.5\% of the total flux density.   Assisted by the boost in WSRT sensitivity  by the availability of a larger bandwidth 
since  April 2002, detailed polarization analysis became possible. These turned out to be very informative as we describe now.  

\begin{figure}
\includegraphics[width=85mm]{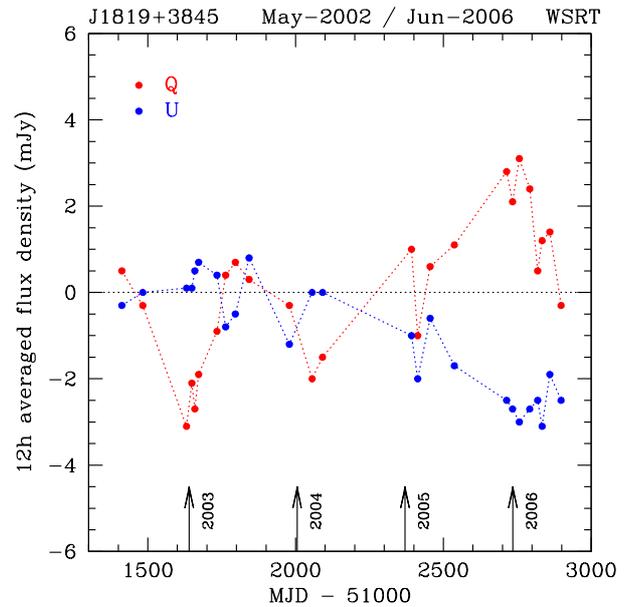}
\caption{Long-term evolution of the flux density of Stokes $Q$ and $U$, for twenty-five, 6\,cm 12~h syntheses taken in the period from May 2002 to June 2006.  The plotted values are averaged over (mostly) 12~h tracks; the peak values in $Q$ and $U$ often reach values of $\pm$ 5-7 mJy.  The uncertainty  on the data points is mostly due to calibration errors: the thermal noise error after 12~h averaging is less than 0.05~mJy. We estimate the uncertainty  on Stokes U  to be typically 0.1\% of the Stokes I flux density or about 0.3 mJy. The uncertainty  on Stokes Q is larger, about 0.25--0.5\% of Stokes I, and this corresponds to about 0.75--1.5 mJy.  These uncertainties  are consistent with the short-term variations between the various epochs.  Detailed lightcurves are shown in other figures.}
\label{fig:qu_evolution}
\end{figure}

Like the total intensity, the polarized emission is variable on a
timescale of tens of minutes.  However, changes in the {\it nature}
of the polarization variability are also observed on a timescale of years (cf.\,Fig. \ref{fig:qu_evolution}),
and we can use this to gain additional insight in the evolution of  structure internal to the 
source, as opposed to scintillation induced apparent variability. From these data we derive 
the apparent speed of internal  source motions.

\begin{figure}
\includegraphics[width=90mm]{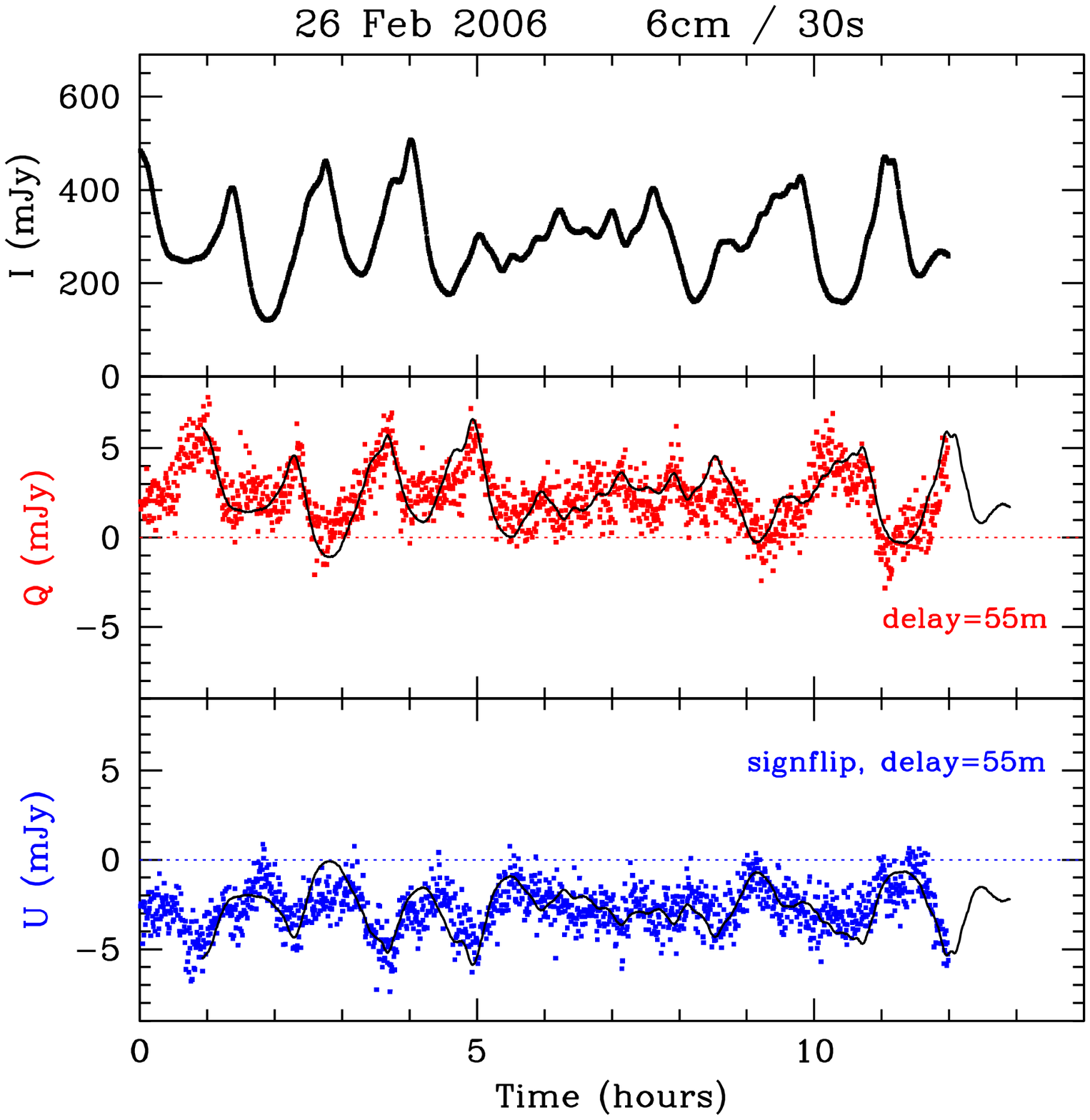}
\caption{A demonstration of the strong correlation between polarized and unpolarized emission.  The polarized emission  lags the unpolarized emission, indicating that the centroid of the polarized emission is angularly offset from the bulk of the unpolarized source emission.  At this particular epoch, 26 Feb 2006, the delay was about 55\,min and similar for both Stokes $Q$ and $U$. A typical noise uncertainty after 30~s integration on the $I$, $Q$ and $U$ data points is about 1.5 mJy. This is less than the width of the lightcurve for Stokes I.  To create a flux density-scaled Stokes $I$ lightcurve, overplotted on the Stokes $Q$ and $U$ lightcurves,  we subtracted the 12~h averaged total flux density, multiplied the difference by  a scale factor, typically 0.015--0.03, and added the average flux of the relevant Stokes parameter as presented in Fig.\ref{fig:qu_evolution}.  The Stokes $U$ signals have  negative polarity on this epoch, hence we flipped the sign of the scaled and shifted Stokes $I$ signals.  In this plot the data were averaged to a timescale of 30\,s.}
\label{fig:polDemo}
\end{figure}

There is a strong cross-correlation between the fluctuations in Stokes parameters $I$, $Q$ and $U$.  Figure\,\ref{fig:polDemo} illustrates the nature of this correlation for a clear-cut and unambiguous case on 26 February 2006.  The peaks in both $Q$ and $U$ are delayed by about 55 minutes from those in Stokes $I$. The systematic nature of the offset --- the delay is the same for all $I$ and $Q$ (or $U$) scintles within each 12~h observation --- indicates that the delay reflects substructure in the emission of the scintillating source.  Such a delay arises naturally if there is an angular displacement between the components housing the polarized and unpolarized emission; a time delay is observed as the structure in the ISM responsible for the scintillation passes first in front of one component and then the other.  The magnitude of the time delay is related to the magnitude of the separation between components.  The correlation between Stokes $I$, $Q$ and $U$ is particularly good in most of the either observations in the  six-month period from December 2005 to Jun 2006. The time delay, however, varies systematically from 240 min on 10 December 2005, 75 min in January 2006, 55 min in February-March 2006, and increases again to 80 min in June 2006. The good correlation  --- with a correlation coefficient typically 60\% or more ---  implies that the polarized emission remains dominated, in that period, by a single polarized feature. We ascribe these components  to the core ($I$) and the tip of the jet ($Q$ and $U$), respectively.  The typical flux ratios  $Q/I$ and $U/I$ are about 1.5--3\%.


Similar time offsets are observed in other IDV sources and are also interpreted in terms of offsets between structures in the source (Bignall et al.\,2003; Rickett et al.\,2002).  Although random refractive gradients in the interstellar medium could cause similar time offsets in principle, they would cause the offset to vary randomly on a timescale comparable to the scintillation timescale, and they could not explain the separation between the linear polarization and the total intensity\footnote{The $80$\,min temporal displacement observed on 22 Jan 2006 is equivalent to an angular separation of $\sim 5 \times 10^{-9} d_{\rm pc}^{-1}\,$rad, where $d_{\rm pc}$ is the distance to the Faraday active region in parsecs.  In a magnetoactive medium an RM gradient $d{\rm RM}/dx$ causes a displacement of the left- and right-hand circularly polarized wavefronts by an angle $(\lambda^3/2\pi) \, d{\rm RM}/dx$ (Macquart \& Melrose 2000), so to cause the angular separation observed here one requires an RM gradient of $4  \times 10^{12} d_{\rm pc}^{-1}\,$rad\,m$^{-2}$\,pc$^{-1}$.  However, even if such an extreme RM gradient were feasible, it would still not cause a displacement of the {\it linear} polarization from the emission in total intensity, since the natural modes of a cold magneto-ionic medium are circularly polarized.}.

We now return to the hourly variations in the linear polarization observed in all 12~h syntheses.  
The variations that are observed in a scintillating source reflect its underlying structure.  The variations in total intensity and polarization variability are the convolution of the point-spread function of the scintillation pattern with the brightness distribution of the source structure (Little \& Hewish 1966; Rickett et al.\,2002; Macquart \& de Bruyn 2007).   

The time delay between any two Stokes parameters $X$ and $Y$ is related to an angular offset, ${\btheta}_{XY}$, in the centroid of the emission within the source by (Coles \& Kaufman 1978),
\begin{eqnarray}
\Delta t_{XY} &=& - \frac{D \btheta_{XY} \cdot {\bf v} + (R^2-1) (D \btheta_{XY} \times \hat{\bf S})({\bf v} \times \hat{\bf S}) }{v^2 + (R^2-1)({\bf v} \times \hat{\bf S})^2}, \\
&\approx& - \frac{D\btheta_{XY} \times \hat{\bf S}}{\vert {\bf v} \times \hat{\bf S} \vert}, \quad R^2 \gg 1,
\end{eqnarray}
where ${\bf v}$ is the velocity of the scintillation pattern at the epoch at which the offset is measured, $D$ is the distance to the scattering screen, $R$ is the anisotropy ratio of the scintillation pattern and $\hat{\bf S} = (\cos \beta, \sin \beta)$ is the direction of its long axis measured with respect to the RA axis.  The best estimates of the scintillation parameters are given in Table \ref{tab:ScintParams}; however, we note that some parameter estimates are subject to degeneracies in the solution for ${\bf v}$.

\subsubsection{Evolution 2003-2006}

If there are multiple polarized features in the jet, possibly with different polarities, the comparison between $I$, $Q$ and $U$ lightcurves can become rather difficult to unravel.   We have investigated all  25 12~h syntheses in the period from April 2002 to June 2006 to look  for cases of unambiguous delays. In about half of the 25 epochs, a unique delay with high correlation coefficient could be derived, especially if there are one or more unusually bright scintels in the Stokes $I$ lightcurve.  In several  cases, e.g. 10 January 2003 (as well as 20 January 2003,  not shown) there is good evidence for the existence of multiple Stokes $U$ components.  In other cases, e.g. 21 February 2004,  the oscillatory behaviour in Stokes $I$ does not permit a unique delay to be determined. 

\begin{figure*}[htbp!] 
 \begin{tabular}{cc} 
\includegraphics[width=85mm]{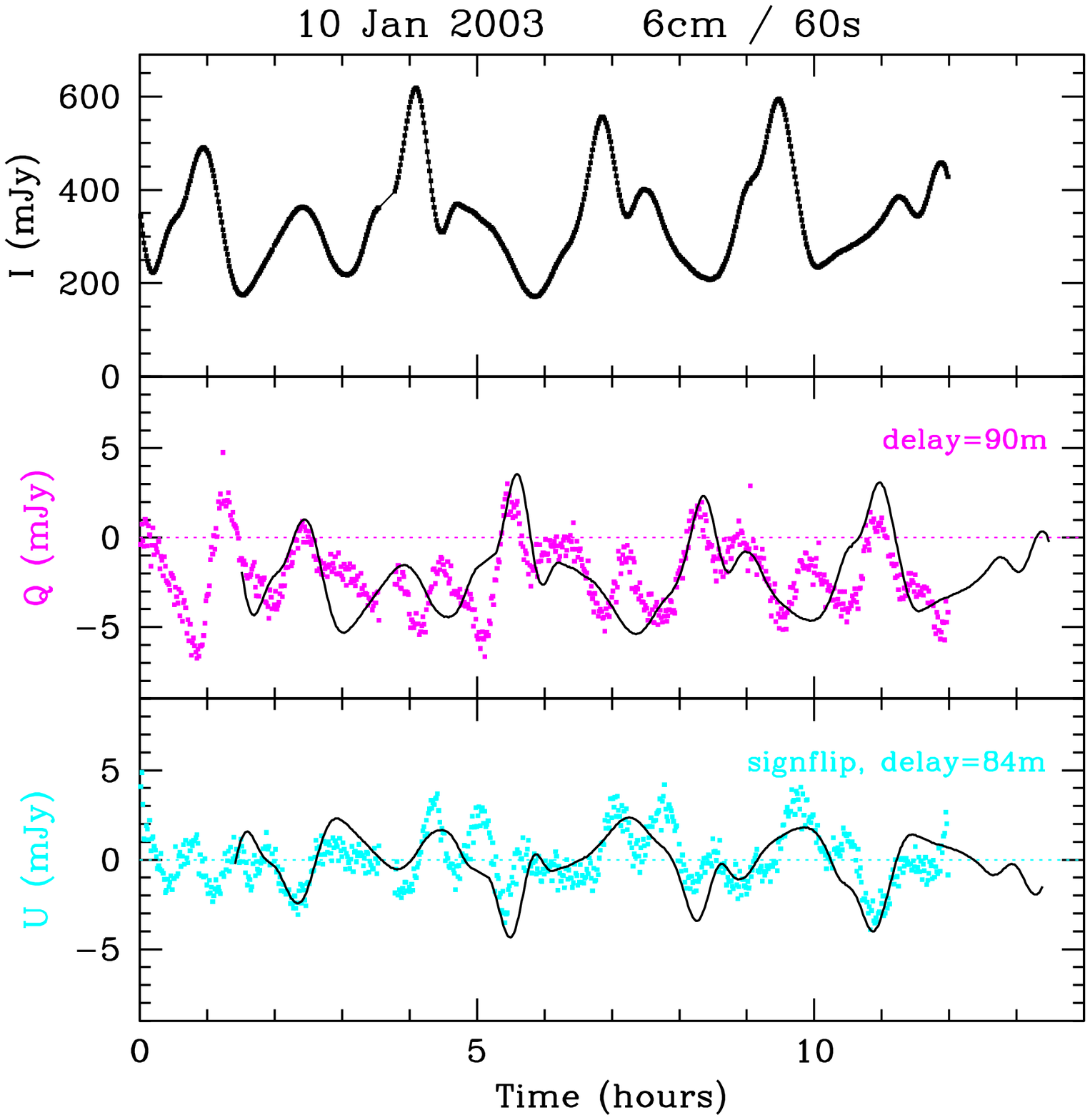} & \includegraphics[width=85mm]{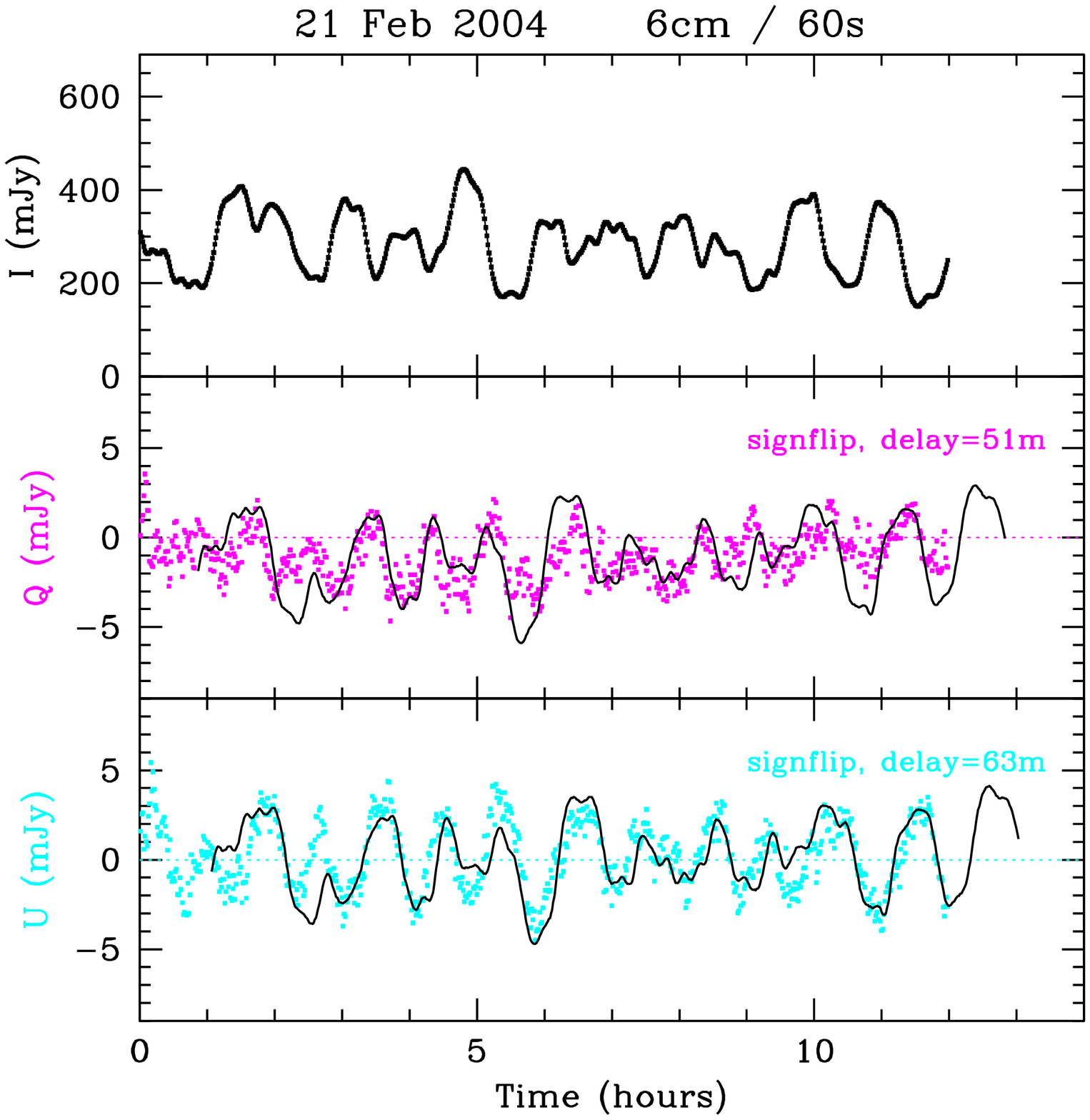} \\
\includegraphics[width=85mm]{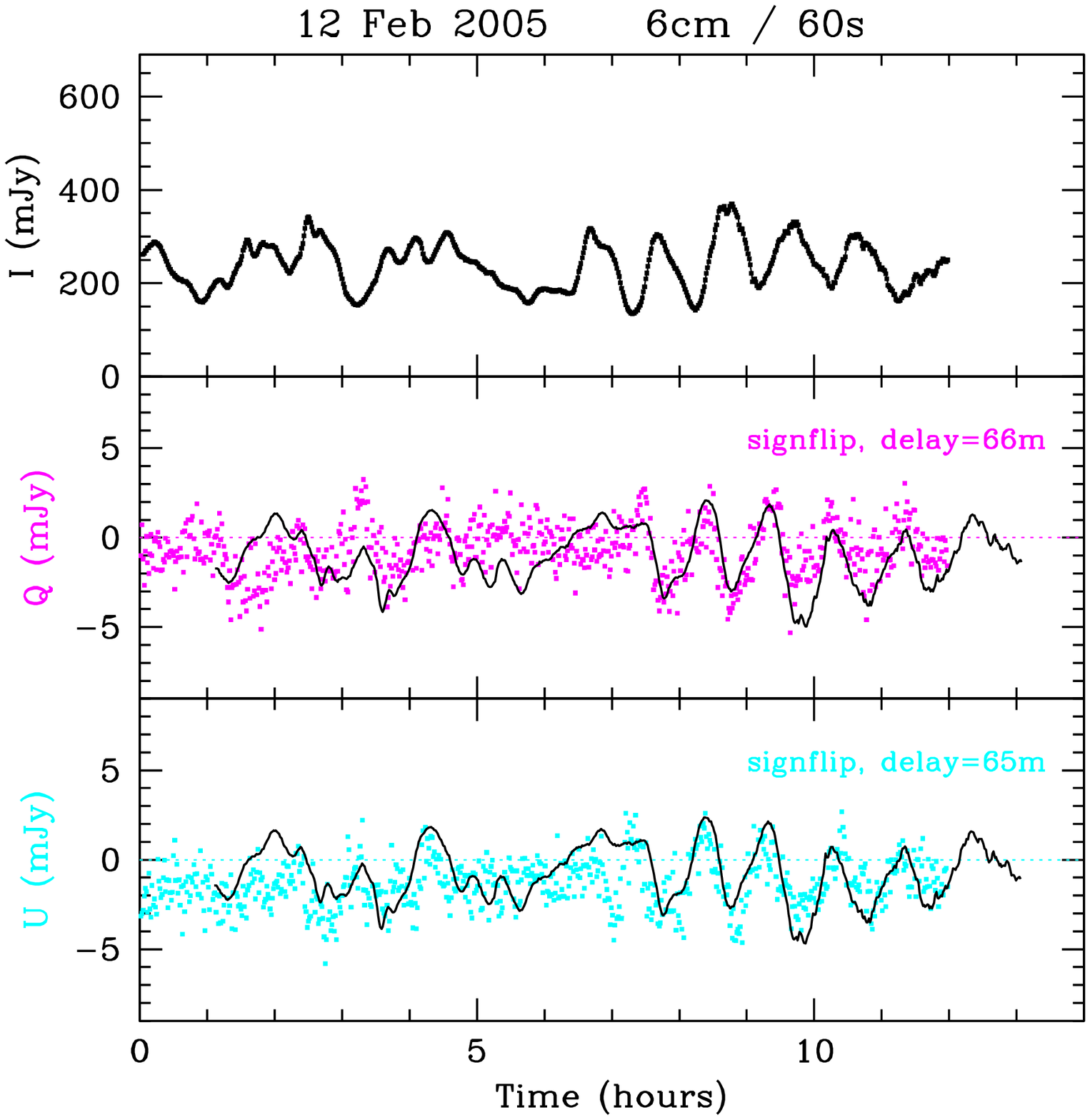} & \includegraphics[width=85mm]{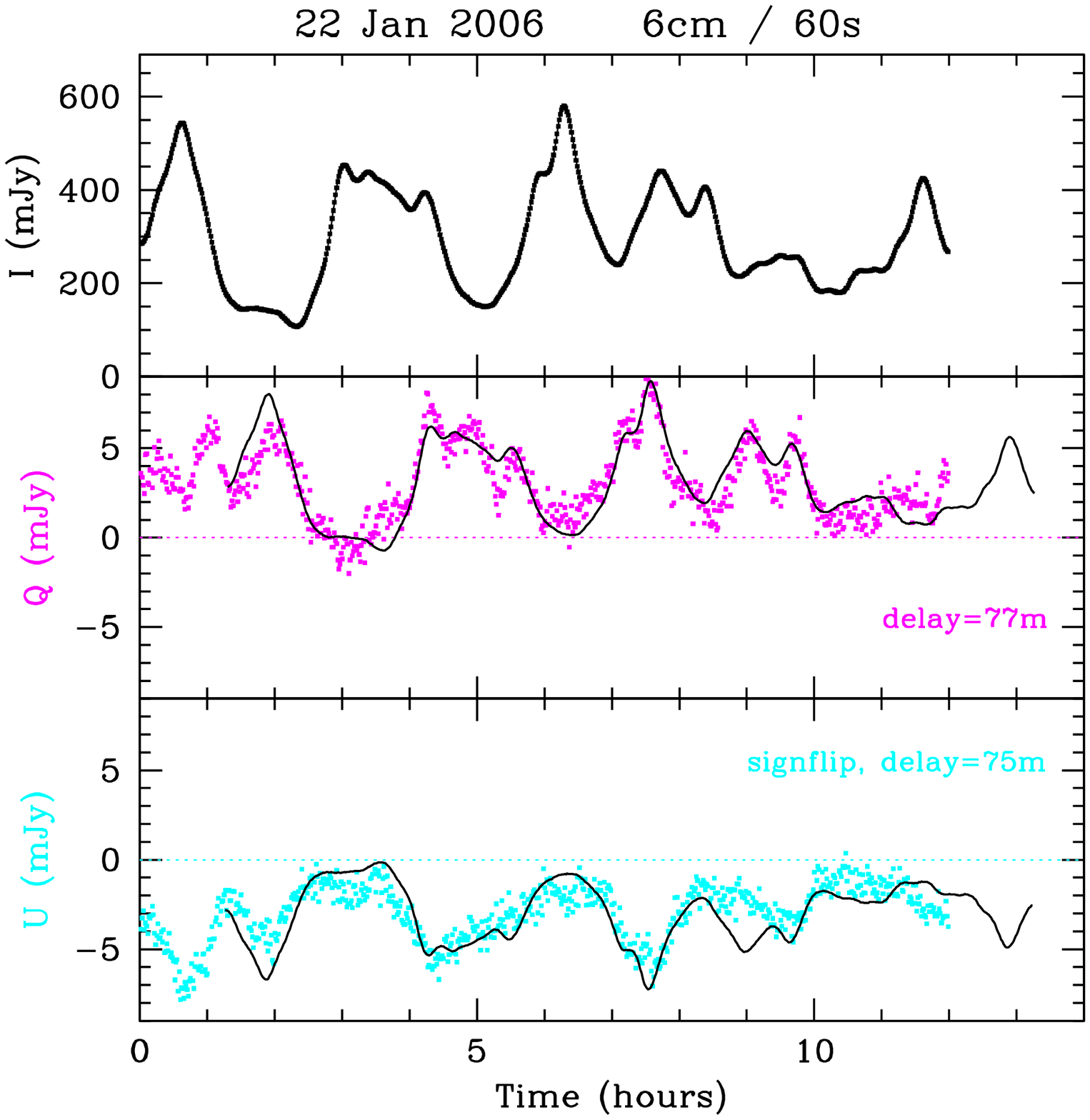} \\
\end{tabular}
 \caption{The variations in $I$, $Q$ and $U$ in J1819$+$3845 over a succession of years.  Each panel shows lightcurves for Stokes $I$, $Q$ and $U$. Typical errors are as described in Fig. 7, but about 1.4$\times$ smaller due to the increased integration time (60~s as against 30~s). Each epoch was chosen to represent a similar stage in the annual cycle of timescale variations exhibited by the source, facilitating meaningful comparison between epochs.  The thin black curves in each of the $Q$ and $U$ panels show a scaled and shifted version of the Stokes $I$ lightcurve for comparison.} \label{fig:PolnMontage} 
 \end{figure*}

Figure \ref{fig:PolnMontage} displays the variability exhibited both in linear polarization and total intensity over four epochs spanning the time interval Jan 2003 to Jan 2006.  The four epochs in Fig.\,\ref{fig:PolnMontage} are chosen so that they fall in the same part of the annual cycle of the variability timescale.  These epochs are quite representative of the behaviour in this part of the year. 
\footnote{ Users that have an interest in further analysis of the data should contact the 1st author; on his webpage (http://www.astron.nl/$\sim$ger/) under the WSRT header we have presented some samples of the data for initial exploration.}
Thus the value ${\bf v}$ is approximately common to all epochs and this removes, to lowest order, ambiguities in the interpretation of $\Delta t_{XY}$ between each of the four datasets due to uncertainties in the change of ${\bf v}$ and the associated angle between ${\bf v}$ and $\hat{\bf S}$. Any change in $\Delta t_{XY}$ between epochs can therefore be attributed mainly to evolution of source structure.  

As noted above, there is a clear one-to-one correspondence between the hourly fluctuations in $I$, $Q$ and $U$ in the 2006 datasets.   However, in the instances where the $Q$ or $U$ intensity changes sign within the 12~h track --- which is the case in many of the 2003 and nearly all of the 2004 and 2005 polarization lightcurves --- the variations cannot be reproduced by a single polarized component; the source must be comprised of at least two components emitting in that Stokes parameter.  For instance, the variations in 2004 in Stokes $U$ could be reproduced with either two scintillating components with opposite signs of $U$, or one scintillating component and a steady or slowly scintillating component with opposite sign.  We note that because the whole source is extremely  compact (less than 1 mas, Moloney 2010), {\it all} parts of the source, including a smooth jet,  are expected to scintillate at detectable levels within a 12~h track in the fast season. The typical timescale for such variations is hours, destroying the one-to-one correlation between $I$, $Q$ and $U$.

In addition to a visual analysis of the $I$, $Q$ and $U$ lightcurves we also performed a  cross-correlation analysis of the lightcurves in Fig.~\ref{fig:PolnMontage}.  This was used to derive the time delays shown in Table \ref{tab:timeDelays}.  In 2003 and 2006 the relationship between the total intensity and linear polarization is  relatively simple, and the time offset between the variations is unambiguous.  However, the behaviour in 2004 and 2005 is more complicated because the cross-correlation function displays a number of peaks of comparable magnitude, indicative of aliasing between the highly periodic fluctuations present in all lightcurves.  If we apply the additional constraints that the sign of the cross-correlation is preserved between epochs and that the $Q-U$ delay is consistent with the $I-Q$ and $I-U$ delays, then the time-delay values indicated by asterisks in Table \ref{tab:timeDelays} represent the most likely delays.  This assumption is subject to the assumption that the intrinsic source polarization does not vary strongly with time.  This is supported by the fact that the polarization amplitude remains comparable between epochs, and thus that the synchrotron opacity likely remains stable.  There also appears to be no appreciable Faraday rotation by a foreground plasma screen within the source.  We have search for the effects of such a screen by examining the data for PA swings across the 160\,MHz observing band and are able to rule out RMs of the magnitude required here to significantly change the PA of the emission.

\begin{figure}
\includegraphics[width=80mm]{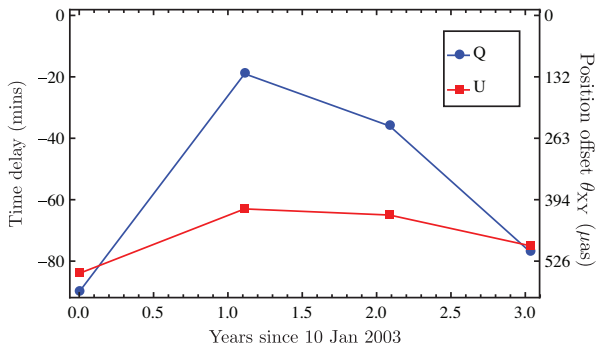}
\caption{A plot of the time delay of the Stokes $Q$ and $U$ emission with respect to the total intensity determined from the four epochs shown in Table \ref{tab:timeDelays}.  The right-hand axis shows the equivalent angular separation in microarcseconds with respect to the unpolarized emission using the relation in eq.\,(\ref{DeltatXY}).} 
\label{fig:polnComponentPlot}
\end{figure}

Figure\,\ref{fig:polnComponentPlot} shows the evolution of the component separations with time according to this solution.  The simplest interpretation is that each of the four datasets represent a snapshot of the source structure that captures the propagation of polarized components moving relative to the unpolarized core along the jet axis of the 
quasar\footnote{The overall framework of our interpretation is corroborated by 
global VLBI observations at frequencies of 8.4, 15 and 22\,GHz in June 2003 and January 2006 presented by Moloney (2010).  These show that the source possesses a barely resolved `core-jet' structure, with the jet oriented towards the North. The total angular extent of the source is less than 1 mas. These observations also reveal faint polarized emission at mJy  levels, located at what presumably is the end of a `jet'.  The WSRT flux density monitoring at 8.4\,GHz had insufficient sensitivity to study polarization. Based on the source spectral index, we expect significant opacity effects between 4.9 and 8.4\, GHz, hence a straightforward comparison between the VLBI and WSRT results is not possible. However, the VLBI observations have been useful to guide the interpretation of the  rapid $I$, $Q$ and $U$ variations due to scintillations. In \S\ref{sec:SrcProp} we return to the VLBI results when we derive the distance of the scattering screen.}.  
It appears straightforward to apply this interpretation to the observations 2004-6, but not to the measured time delay in 2003.  The 2003 delay would only be consistent if the 2004-6 data track a different ejection event relative to the earlier epoch.

\subsubsection{Component separation speeds}
Notwithstanding the difficulties of identifying specific offsets uniquely with certain polarized components, it is still possible to determine an order-of-magnitude estimate for the velocity of polarized components within the jet.  One can relate these time offsets to physical offsets within the source, $\btheta_{XY}$, by adopting the solution for the velocity vector ${\bf v}$ on 30 Jan determined by the parameters in Table \ref{tab:ScintParams} and using the approximation $R^2 \gg 1$ applicable to the scintillations of J1819$+$3845:
\begin{eqnarray}
\Delta t_{XY} = 76 \left( \frac{D_{\rm scr}}{2\,{\rm pc}} \right) \left( \frac{\theta_{XY}}{500\,\mu{\rm as}} \right) \sin (\zeta  - 187 \pi/180 ) \,\,{\rm min}, \label{DeltatXY}
\end{eqnarray}
where $\zeta$ is the angle between the direction of $\btheta_{XY}$ and the RA axis. Over the period 2004-6, the average component separation speed is 13$\,\mu$as\,yr$^{-1}$ in $Q$ and 4$\,\mu$as\,yr$^{-1}$ in $U$.  At the source redshift of $z=0.54$, 1$\,\mu$as is equivalent to 0.0064\,pc, and this translates to apparent component speeds of $0.29\,c$ in $Q$ and $0.08\,c$ in $U$.  Ambiguities in the unique identification of polarized components between epochs mean that these speed estimates may be incorrect in detail, but they are correct to within an order of magnitude, which is sufficient for the present purpose of comparing typical component speeds to the expansion speed necessary to cause the source to have stopped scintillating in 2006-7.

This estimate can be compared against the results of power spectral modelling of the total intensity emission of the source, from which an expansion speed $3.4 \pm 0.3\,c$ was derived between the extremely compact components that appeared in the period 2004-2005 (Macquart \& de Bruyn 2007).   It is not completely clear how the structure in the total intensity relates to the polarization structure; a detailed cross-spectral analysis of $I$, $Q$ and $U$ is beyond the scope of this paper.  It is nonetheless clear that the typical speed of component separations is well short of the $>10\,c$ required {\it across the entire source} to cause the scintillations to vanish during 2006--7.  This, combined with the fact that VLBI measurements indicate that there has been no source size increase commensurate with the cessation of scintillation, and the source spectrum and flux density have not changed markedly, makes it implausible to relate this event to changes within the source. This conclusion can probably be extended to the last year of rapid scintillations. The systematic change in the time delay between $I$, $Q$ and $U$ features  in the period Dec 2005 till June 2006, as discussed above, are commensurate with the changes in projected velocity during that period; they therefore are consistent with a picture that the `tip of the jet' in 2006 did not appear to move substantially.

\begin{table}
\begin{center}
\begin{tabular}{|c|c|}
\hline
Parameter & value   \\ \hline
D	& $1-3$\, pc \\
SM & $(2.5 \pm 0.4) \times 10^{17}\,$m$^{-17/3}$ \\
\null & $=(8 \pm 1) \times 10^{-3}  \,$m$^{-20/3}$\,kpc\\
$v_\alpha$ & $33.5$\,km\,s$^{-1}$ \\
$v_\delta$ & -$13.5$\,km\,s$^{-1}$ \\
R & $14_{-8}^{+>30}$  \\
$\beta$ & $83^{\circ} \pm 4^{\circ}$ (N through E) \\ \hline
\end{tabular}
\end{center}
\caption{Parameters associated with the interstellar scintillation of J1819$+$3845 collated from Dennett-Thorpe \& de Bruyn (2003) and Macquart \& de Bruyn (2007).  The parameters $v_\alpha$ and $v_\delta$ refer to the velocity of the scattering screen with respect to the heliocentre.  Its velocity with respect to Earth changes during the course of the year.  On 30 Jan the velocity with respect to Earth is $(v_{\oplus_{\alpha}},v_{\oplus_{\delta}}) = (9.2,-31.9)\,$km\,s$^{-1}$.} 
\label{tab:ScintParams}
\end{table}

\begin{table*}
\begin{center}
\begin{tabular}{|c|c|}
\hline
\begin{tabular}{ccc}
\multicolumn{3}{c}{10 Jan 2003} \\ \hline
Stokes pair & $\Delta t$ (min) & $\rho$ \\ \hline
$I-Q$ & -90 & 0.55\\
$I-U$ & -84$^*$/-14 & -0.63$^*$/ 0.53\\
$Q-U$ & -2$^*$/-89& -0.41$^*$/0.56\\
\end{tabular}
 & 
\begin{tabular}{ccc}
\multicolumn{3}{c}{21 Feb 2004} \\ \hline
Stokes pair & $\Delta t$ (min) & $\rho$ \\ \hline
$I-Q$ & -51/-19$^*$ & -0.52/0.44$^*$\\
$I-U$ &  -63$^*$/-30& -0.71$^*$/0.72 \\
$Q-U$ & -39$^*$/-11 & -0.59$^*$/0.56 \\
\end{tabular}  \\
\hline
\begin{tabular}{ccc}
\multicolumn{3}{c}{12 Feb 2005} \\ \hline
Stokes pair & $\Delta t$ (min) & $\rho$ \\ \hline
$I-Q$ & -66/-36$^*$/-5& -0.41/0.50$^*$/-0.49 \\
$I-U$ & -65$^*$/-36/-4 & -0.48$^*$/0.42/0.40\\
$Q-U$ & -61/-29$^*$/2/29 & 0.39/-0.26$^*$/0.39/-0.42$^*$\\
\end{tabular}
&
 \begin{tabular}{ccc}
\multicolumn{3}{c}{22 Jan 2006} \\ \hline
Stokes pair & $\Delta t$ (min) & $\rho$ \\ \hline
$I-Q$ & -77 & 0.89 \\
$I-U$ & -75 & -0.65\\
$Q-U$ & 6 & -0.60\\
\end{tabular}  \\
\hline
\end{tabular}
\end{center}
\caption{The time delays between the variations in the various Stokes 
parameters for the four epochs shown in Fig.\ref{fig:PolnMontage}.  In
the 21 Feb 2004 and 12 Feb 2005 datasets there are multiple peaks with
similar  correlation coefficients, $\rho$, and in these cases we
report the delay and correlation coefficient associated with each
peak.   A negative delay, $\Delta t$, between Stokes parameters $X$ 
and $Y$ means that the variations in Stokes parameter $Y$ lag those 
in $X$.  The time-delay values indicated by asterisks represent the most likely delays subject to the additional constraint that the sign of the cross-correlation is preserved between epochs, and that the $Q-U$ delay is consistent with the $I-Q$ and $I-U$ delays.  We estimate the error associated with each measurement of $\Delta t$ to be $\approx 2\,$min (1$\sigma$).  By far the largest source of error in the time lag relates to the ambiguities introduced by aliasing.}
\label{tab:timeDelays}
\end{table*}

\section{Screen properties} \label{sec:SrcProp}


Given that J1819$+$3845 has not exhibited internal structural changes capable of causing the cessation of scintillations observed in 2006-7, the phenomenon must be attributed to a change in the turbulent scattering screen in front of the source.  Here we infer the properties of this scattering region.

\subsection{Distance of the screen}

\begin{figure}
\includegraphics[width=85mm]{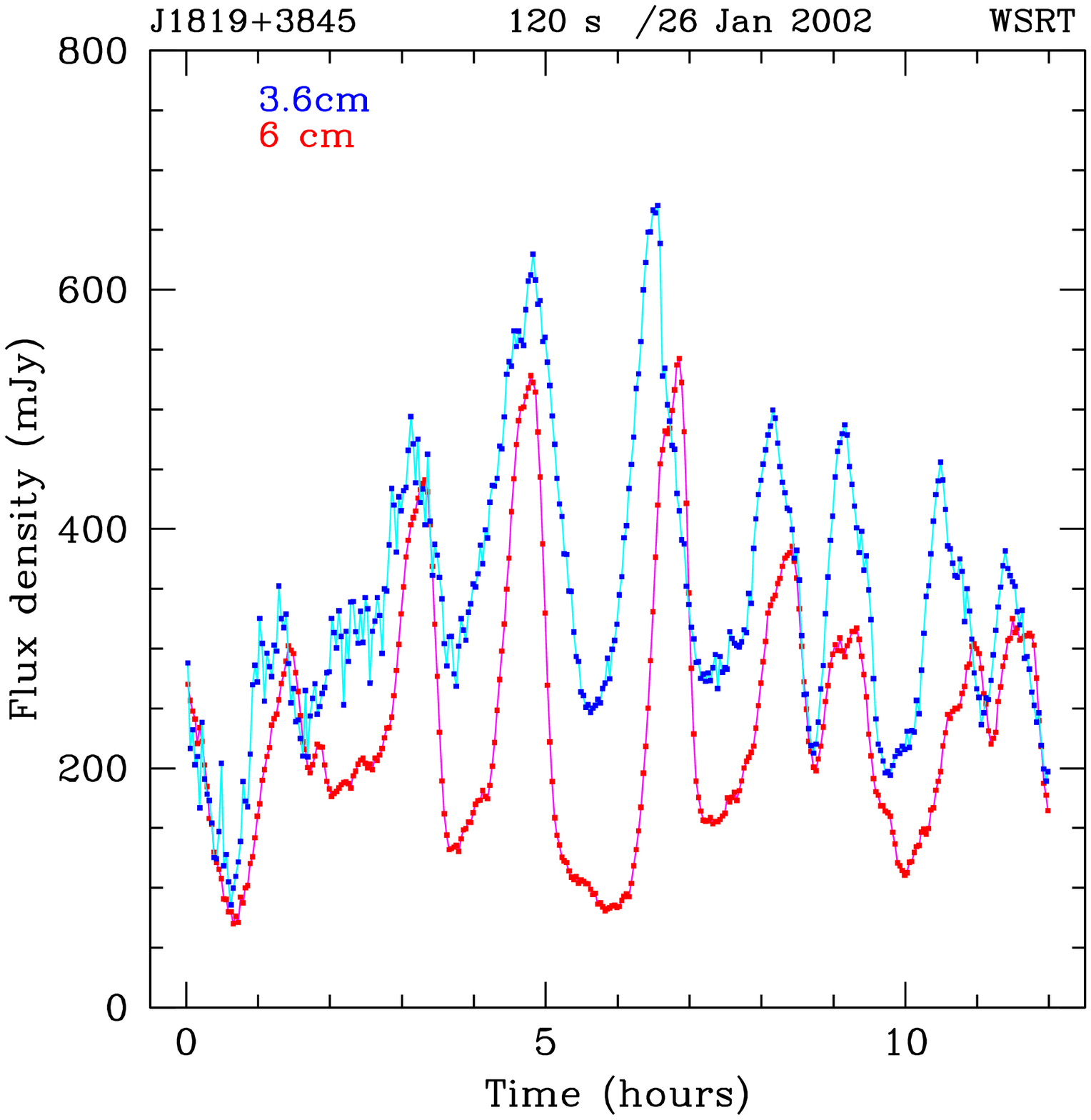}
\caption{The 3.6\,cm and 6\,cm lightcurves of J1819$+$3845 on 26 Jan 2002. Typical uncertainties on the 120~s averages are about 4 mJy for 3.6cm and 1 mJy  for 6cm. This does not include systematic flux calibration errors, which we estimate at 4\% for 3.6cm and 2\% for 6~cm. There appears to be a systematic time delay of about 10 min between the peaks of the scintles, with the 6\,cm scintles lagging. In addition there appears to be a systematic asymmetry in the profiles near the peak, similar to that seen on many other occasions, see e.g.\,Fig.\,\ref{fig:3.6cmdouble}.}
\label{fig:3on6}
\end{figure}

\begin{figure}
\includegraphics[width=85mm]{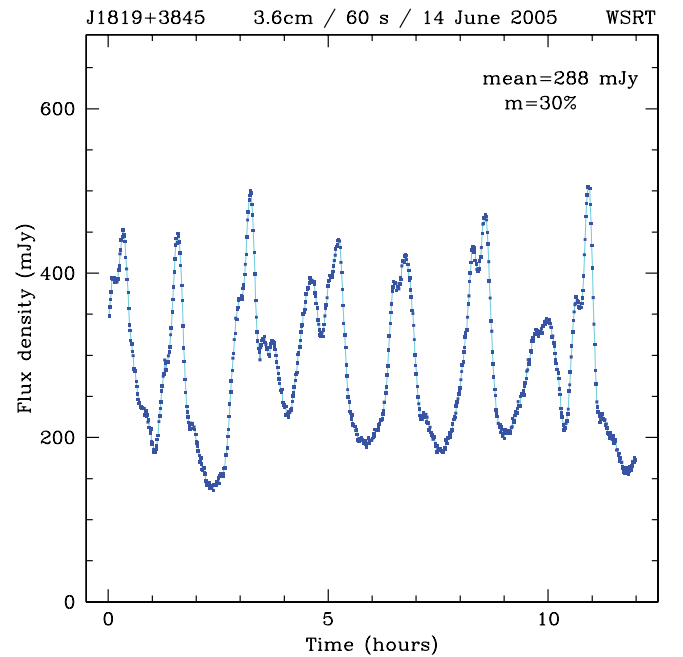}
\caption{The lightcurve of J1819$+$3845 at 3.6\,cm on 14 June 2005 reveals time-resolved double peaks, with a separation of about 15 minutes between features in the lightcurve near the peaks of most scintles. The uncertainty  on individual 60~s datapoints is about 6 mJy.}
\label{fig:3.6cmdouble}
\end{figure}

The rapid scintillations in J1819$+$3845 have been associated with a region
of turbulent material within 1-3\,pc of Earth that moves relative to
the line of sight  with a best-fit scintillation velocity of
$(v_\alpha,v_\delta)=(-33.5,13.5)\,$km\,s$^{-1}$ (Dennett-Thorpe \& de
Bruyn 2003).  There is some degeneracy in the solution for ${\bf v}_{\rm ISS}$, and we henceforth adopt $v_{\rm ISS} = 35\,$km\,s$^{-1}$ relative to the barycentre of the Solar System.
Modelling of the power spectrum of intensity variations at 6\,cm
(Macquart \& de Bruyn 2007) yields a distance of $2.0 \pm 0.3\,$pc,
for this scintillation velocity.  However an independent and more robust determination
of the screen distance is possible by comparing the temporal shifts
between the total intensity and polarization variations, of the sort
discussed in \S\ref{subsec:polnComponents}, with the angular separation 
of these components as deduced from high-resolution VLBI  observations.
Global VLBI data taken for this purpose were obtained in June 2003 
and have been described in Moloney (2010). They reveal that the polarized emission at 8.4\,GHz is 
displaced from the total intensity peak by about 600 $\pm$ 100 $\mu$as to the North.  The gaps in the 8.4\,GHz VLBI lightcurves and the largely oscillatory nature of the variations during the VLBI observations did not permit a reliable time delay to be determined from the data.  We therefore use the 4.9\,GHz time delays determined from the regular WSRT monitoring.  Before we can relate this {\it angular separation} at 8.4\,GHz to the {\it linear scale} determined from the 4.9\,GHz time delay, we still need to correct for a small frequency-dependent source centroid shift due to synchrotron opacity effects (the 4.9\,GHz emission of J1819$+$3845 was significantly  self-absorbed in the period from 2000 to 2006, see Fig.~\ref{fig:4freqEvolution}). Such  shifts are often seen in astrometric phase-referenced VLBI observations of compact AGN (see e.g. Bignall et al.\,2012).  Many epochs of simultaneous WSRT 4.9\,GHz and 8.4\,GHz observations indeed show that in the fast season (January-June) the scintillation peaks  at 8.4\,GHz arrive, on average, about ten minutes ahead of the peaks at 4.9\,GHz (an early example is shown in Fig.\,\ref{fig:3on6}).  There are also periods when the 8.4\,GHz peaks themselves are clearly double (see Fig.~\ref{fig:3.6cmdouble}) with a time separation of about 15 minutes. These results probably signify 8.4\,GHz structure near the true core.  

At the epoch of the VLBI observations in June 2003 we estimate the time delay between the total and polarized intensity peaks at 4.9\,GHz to be about 50 minutes. Augmenting this with the offset of 15 minutes between the core at 3.6\,cm and 6\,cm, we estimate a temporal shift of 50 + 15 = 65 minutes between the 8.4\,GHz total and polarized intensity emission regions in June 2003.   Multiplying this time delay by the 35 km\,s$^{-1}$ N-S velocity of the quasar relative to the screen at that time (see Dennett-Thorpe \& de Bruyn, 2003), we deduce a linear separation of the scintles at the Earth of $1.4 \times 10 ^{10}\,$cm. Combining this with the angular displacement of 600\,$\mu$as then yields a distance to the screen of $5 \times 10^{18}\,$cm, or 1.5\,pc. We estimate this direct determination to be in error by at most 0.5\,pc.  We therefore find excellent agreement with the distance estimate given above as deduced from the rapid 6\,cm scintillations in 2004 and 2005 (Macquart \& de Bruyn 2007).   

\subsection{Anisotropy}

The observed spatial scintillation pattern in J1819$+$3845 is extremely anisotropic, and this is often taken as evidence that the turbulence in the scattering region must be similarly anisotropic.  The scintillation pattern is anisotropic by a factor of at least 6:1, with a best fit value of 14:1 (Dennett-Thorpe \& de Bruyn 2003).  The observed scintillation pattern is a convolution of the point-spread function of the interstellar turbulence (i.e.\,the pattern that a point-like source would exhibit through the same scattering medium) with the source structure.  Anisotropy, therefore, can either reflect elongation in the underlying source structure --- which is plausible if the source exhibits a jet-like morphology, as suggested by the analysis in \S\ref{subsec:polnComponents}  --- or it could reflect anisotropy in the underlying power spectrum of the turbulence responsible for the scintillations.  The  latter explanation is also highly plausible, since pulsar speckle-imaging also reveals anisotropic scattering in the ISM, with elongations exceeding 40:1 (Brisken et al.\,2010).  In fact, it is possible to reproduce the variability characteristics of J1819$+$3845 with an infinitely anisotropic, line-like, scintillation pattern (Walker et al. 2009).  However, since the anisotropy is deduced primarily on the basis of the annual cycle in the variability timescale, it is impossible to attribute this to the turbulence-based anisotropy with certainty.  Models of the power spectrum of the variations  (Macquart \& de Bruyn 2007), in which the two kinds of anisotropy can be partially separated, do not {\it require} the power spectrum of the electron density fluctuations to have an anisotropy ratio exceeding 3:1.

However, the N-S elongation of the core-jet structure of the source is at variance with orientation of the the anisotropy.  In fact, the scintillation pattern is elongated nearly orthogonal to the jet axis, and this argues that the anisotropy reflects the characteristics of the interstellar turbulence responsible for the scintillations.

\subsection{Screen extent}
Several properties of the turbulent region associated with the scintillation can be estimated based upon the manner in which the scintillations ceased.  The cessation of scintillations in J1819$+$345 occurred on a timescale of less than 257 days.  At a speed of 35\,km\,s$^{-1}$ across the line of sight, this requires the edge of the scattering region to possess a thickness of less than 5.2\,AU.
The fact that scintillations were observed over the period 1999-2007 places a lower limit on the extent of the scattering screen in the direction of the transverse velocity of 59\,AU, which means that the screen should subtend an angular scale exceeding 1\arcmin on the sky for a screen distance of $1\,$pc.  This represents a lower limit in the sense that it is probable that the source exhibited fast scintillations prior to its first detection in 1999.  



\begin{figure}
\includegraphics[width=85mm]{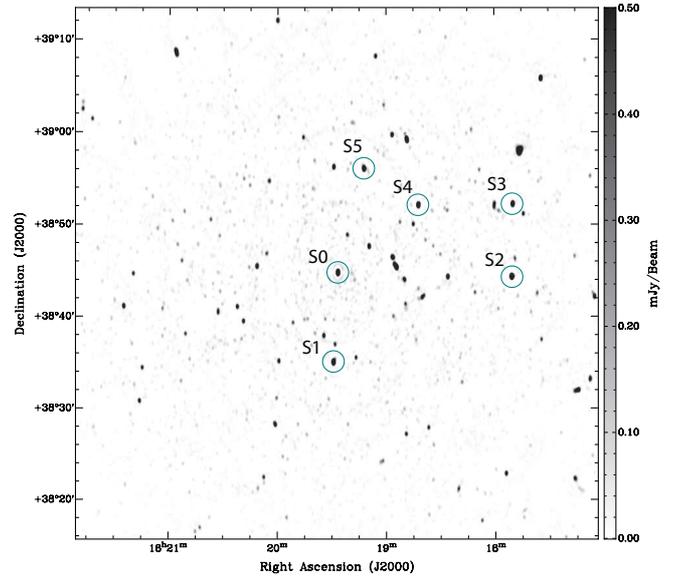}
\caption{Grey-scale image of the field surrounding J1819$+$3845 at 21\,cm. J1819$+$3845 is the source in the centre ($S0$), sources labelled $S2$ and $S3$ are the variable flat spectrum sources, and sources $S1$, $S4$ and $S5$ are steep spectrum sources. For other salient properties of this image we refer to the text.} 
\label{fig:21cmFieldImage}
\end{figure}

Given the large extent of the screen, it is natural to consider whether any other compact sources in the vicinity of J1819$+$3845 are covered by the same turbulent screen and exhibit a similar degree of variability.  Likely candidates probably should have a flat spectrum indicating compactness. We have examined the brighter sources in the many 21\,cm images of the field for variability. Two relatively nearby sources have flat spectra between 6 and 85\,cm, suggesting they are compact. To indicate the location of these background sources relative to J1819$+$3845 we show in Fig.\ref{fig:21cmFieldImage} one of the many 21~cm images of the field. Sources S2 and S3 are the brightest sources located about 20$\arcmin$ West of J1819$+$3845. More than 200 other background sources can be seen in this very deep image, which has a noise level of 15 $\mu$Jy and an angular resolution of about 20$^{\prime\prime}$. The field of view is limited by the primary beam of the WSRT, which is 0.6$^{\circ}$ at 21~cm. We refer to Macquart and de Bruyn (2006) for a description of how this image was produced.  Three other bright field sources, designated $S1$, $S4$ and $S5$, have a steep spectrum.  The positions of the five sources are listed in Table \ref{CompactTable}.

%
\begin{table}
\caption{The J2000 positions of the compact sources in the vicinity of J1819$+$3845.  These positions are accurate to 0.5$\null^{\prime \prime}$.} \label{CompactTable}
\centering
\begin{tabular} {ccc}
\hline\hline
Source Label & RA (h m s) & Dec ($^\circ$ $\null^\prime$ $\null^{\prime \prime}$) \\
\hline
J1819$+$3845 (S0) & 18 19 26.55& $+$ 38 45 01.8\\
S1 & 18 19 28.99 & $+$ 38 35 17.9 \\ 
S2 & 18 17 50.42 & $+$ 38 44 30.1 \\
S3 & 18 17 49.76 & $+$ 38 52 22.1 \\
S4 & 18 18 42.00 & $+$ 38 52 20.5 \\
S5 &18 19 12.06 & $+$ 38 56 16.5 \\
\hline
\end{tabular}
\end{table}

\begin{figure}
\includegraphics[width=90mm]{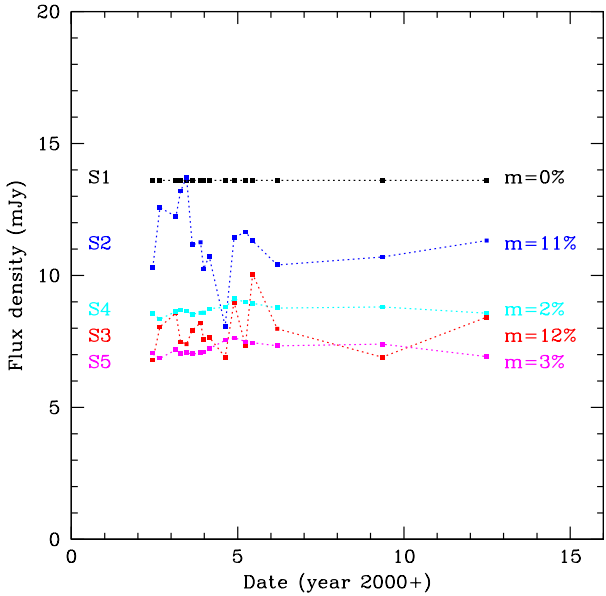}
\caption{Lightcurves at 21\,cm of the five brightest field sources in the 21\,cm image shown in Fig.~\ref{fig:21cmFieldImage}.  The steep spectrum source $S1$ was used for flux normalization. The modulation indices of the sources are given on the right. The typical uncertainty on the flux densities for the various epochs is dominated by systematic errors, which are estimated to be about 2\%. This implies that no believable variations have been detected in sources $S4$ and $S5$.  All fluxes given are apparent fluxes, uncorrected for the primary beam.}  
\label{fig:21cmFieldVariables}
\end{figure}

The lightcurves of these five sources, which are the brightest sources in the field next to J1819$+$3845, are shown in Fig.\,\ref{fig:21cmFieldVariables}.  Sources $S2$ and $S3$  show significant long-term variations with modulation indices of 11\% and 12\%, respectively.  Up to 40\% variations are seen in observations only two months apart in both $S2$ and $S3$. The sources $S2$ and $S3$ are too faint at 21\,cm to detect intra-day variations. The weak 2--3\% variations apparently exhibited by the steep spectrum sources $S4$ and $S5$, however, appear to be correlated. The simplest explanation is that source $S1$, which was used for the normalization of the lightcurves, is perhaps itself slightly variable at a level of about 3\%. Even if true, this does not affect the magnitude of the modulation indices deduced for sources $S2$ and $S3$.    The long-term 21\,cm variations in $S2$ and $S3$  are still a factor 2--3 lower than those shown by J1819$+$3845 in the same time span.  These variations may well be due to turbulence in much more distant material in the interstellar medium. 

\begin{figure}
\includegraphics[width=85mm]{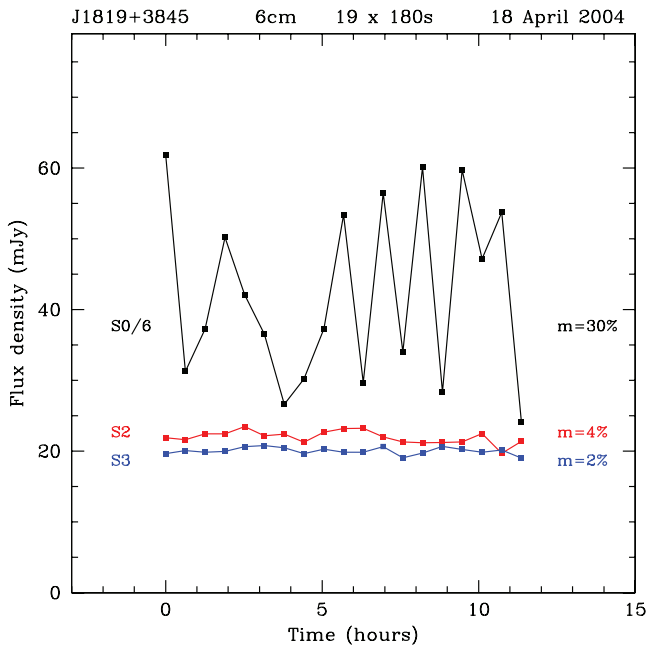}
\caption{Lightcurves at 6cm of the sources $S2$ and $S3$ on 18 Apr 2004  shown with a scaled-down lightcurve  of J1819$+$3845 (S0/6). The observed modulation indices for sources $S2$ and $S3$ shown on the right are consistent with a constant flux density during 12~h.  By contrast, the observed variations in J1819$+$3845, with has a modulation index of 30\%,  are typical for the source in the fast season.} 
\label{fig:FastFieldVars}
\end{figure}

We have also searched for hourly variations at 6\,cm in sources $S2$ and $S3$. 
For these observations, taken on 18 April 2004 in the fast season,  we repointed the WSRT with a duty cycle of 40 minutes between ten flat spectrum sources, including J1819$+$3845, selected from an area of several degrees around J1819$+$3845.  The flux densities of the seven brightest sources ranged from 16--35 mJy; none showed detectable variations on hourly timescales.  
In Fig.~\ref{fig:FastFieldVars} we show the lightcurves of J1819$+$3845, $S2$ and $S3$. 
The contrast between the stable lightcurves of $S2$ and $S3$ on the one hand, and J1819$+$3845 on the other, is dramatic.  The absence of strong scintillations similar to those exhibited by J1819$+$3845 in other compact field sources suggests that either the screen is very small or patchy or that J1819$+$3845 is abnormally compact relative to other flat- and inverted-spectrum sources.  The latter explanation is unlikely given that the brightness temperature of the source is not abnormally high (Dennett-Thorpe \& de Bruyn 2003; Macquart \& de Bruyn 2007).  Our continued monitoring of this field might reveal further constraints on the geometry of the scattering screen.

\subsection{Physical properties}

The scattering region that causes the scintillations in J1819$+$3845 is exceptionally turbulent.
The strength of the scattering associated with a turbulent region is derived using the amplitude of the scintillations and the distance to the scattering region.  For the amplitude of the scintillations observed at 6\,cm and with such a nearby screen distance, one requires $C_N^2 > 0.7 \Delta L_{\rm pc}^{-1}\,$m$^{-20/3}$, where $\Delta L_{\rm pc}$ is the region thickness in parsecs (Macquart \& de Bruyn 2007), and it is likely, for reasons discussed below, that the region thickness is a very small fraction of $1\,$pc.  By comparison, values of $C_N^2 \sim 10^{-3}\,$m$^{-20/3}$ are typically deduced from the scintillations of pulsars, though these estimates are subject to the assumption that the scattering material is distributed homogeneously along the line of sight, which is likely to be a poor approximation in many cases.

The large scattering measure of the turbulent region deduced from the scintillations of J1819$+$3845 places additional severe constraints upon the nature of the scattering medium.  The
scattering measure inferred from the 6\,cm scintillations is $(8.4 \pm 0.6) \times 10^{-3}$\,kpc\,m$^{-20/3}$ (Macquart \& de Bruyn 2007), which corresponds to an increment in the
emission measure ($\rm EM$) of $4.58 \, l_0^{2/3} \, (1+\epsilon^2)/\epsilon^2\,$pc\,cm$^{-6}$ (NE2001 eq. 14), where $l_0$ is the outer scale of the turbulence in AU and $\epsilon = \langle  (n_e - \bar{n_e})^2 \rangle^{1/2} /  \langle n_e \rangle$ is the rms of electron density fluctuations relative to the mean electron density.  This is comparable to the value derived from the H$_\alpha$ intensity towards this source 
from the Virginia Tech Spectral Survey of 0.70 Rayleighs, which implies ${\rm EM} = 2.0 T_4^{0.9}\,$pc\,cm$^{-6}$ (eq.\,1 of Haffner et al. 1998), where $T_4$ is the temperature of the emitting gas in units of $10^4\,$K.  This emission is presumably  dominated by thermal plasma at much larger distances and therefore represents an upper limit to the EM of the nearby screen.  

Since the estimated scattering screen size is at least as large as the resolution element in the VTSS H-alpha spectral line survey, beam dilution is not a plausible explanation for the apparent discrepancy between the measured EM increment and that inferred on the basis of the SM.   The resolution of this discrepancy therefore lies in one of two possible alternatives. (i) The outer scale of the turbulence may not be comparable to the transverse dimensions of the cloud.  It is difficult to speculate on the plausibility of this possibility, since the physical origin of the turbulence is unknown; a model based on anisotropic Kolmogorov turbulence clearly explains the scintillation properties on very small length scales internal to the cloud, but we do not know the spectrum of the turbulence on scales comparable to the cloud size.  We know the overall cloud does not fit well in the paradigm of the Galactic turbulence power spectrum (i.e. there is a disconnect between the cloud and its internal turbulence).  (ii) The parameter $\epsilon$ could be considerably greater than unity.  The available data renders it difficult to distinguish between the likelihood of these two alternatives as the resolution of the apparent inconsistency.

It is possible to estimate the density internal to the scattering region based on estimates of the screen thickness.  One has ${\rm EM}=\int n_e^2 dl = \int \langle n_e \rangle^2 (1 + \epsilon^2) dl$; so, for a medium of thickness $\Delta L$ whose mean and rms density are constant throughout its depth, the mean electron density is 
\begin{eqnarray}
\langle n_e \rangle = \sqrt{\frac{{\rm EM}}{(1+\epsilon^2) \Delta L}}.
\end{eqnarray}
Given the EM calculated on the basis of the scattering measure above, we can estimate the electron density subject to an estimate of the depth of the scattering region.  To this end, we first we note that the optical properties of the scintillation are well-modelled using the thin-screen approximation often used in scintillation theory (Dennett-Thorpe \& de Bruyn 2003; Macquart \& de Bruyn 2007), so from an optics point of view the screen depth cannot exceed more than $\sim10$\% of its distance, i.e. 0.1--0.2\,pc.  
A more stringent constraint, however, is derived from transverse extent of the cloud, which we estimate above to be $59\,$AU or larger.  Assuming that we are not observing the scattering region from a special viewpoint, the depth of the region should be comparable to its transverse extent.  Thus the emission measure places a constraint on the ambient electron density inside the scattering region of 
\begin{eqnarray}
\langle n_e \rangle = 97 \, l_0^{1/3} \epsilon^{-1}  \left( \frac{\Delta L}{100\,{\rm AU}} \right)^{-1/2} \,\,\, {\rm cm}^{-3},  \label{DensityEstimate}
\end{eqnarray}
where we normalize to a fiducial screen depth of $100\,$AU, based upon the arguments outlined above.  However, we note that even if the screen thickness is as large as 0.1\,pc, the implied ambient density of $6 \, l_0^{1/3} \epsilon^{-1}\,$cm$^{-3}$ still exceeds the average density in the warm ionized ISM by over two orders of magnitude\footnote{One reason to consider such large depths is the alternative viewpoint that one of the reasons this scattering region is peculiar is that it may relate to a special viewing geometry.  In this case the scattering region may be highly aspherical, so the screen thickness $\Delta L$ may be much larger than the transverse extent.  This would be the case if the scattering region consisted of a thin sheet viewed nearly edge-on (see the suggestions along these lines by Goldreich \& Sridhar 2006 and Pen \& King 2011) or, in an even more special geometry, if it consisted of a thin cylindrical filament.}.  

The overdensity problem may be partially alleviated by appealing to a small turbulent outer scale (i.e. $l_0$ much less than 1\,AU).  However, this cannot account for the large discrepancy.  Observations of the variability of J1819$+$3845 over 12-hour durations, which probe length scales in the scattering region of 0.01\,AU, show no evidence of a turnover associated with an outer scale (Macquart \& de Bruyn 2007).  This limit suggests that a small outer scale can reduce the above density estimates by only as much as a factor of $4.6$.

The dispersion measure associated with the scattering region follows from the estimate of the mean density:
\begin{eqnarray}
{\rm DM} = 4.7 \, l_0^{1/3} \epsilon^{-1}  \left( \frac{\Delta L}{100\,{\rm AU}} \right)^{1/2} \,\,\, {\rm pc\,cm}^{-3}. \label{DMEstimate}
\end{eqnarray}
Such an increment should be easily measurable if the scattering region were to move across the line of sight of a pulsar. 

Extreme densities such as those derived here are also inferred for the clouds implicated in Extreme Scattering Events (ESEs; Fiedler et al. 1987; Romani et al. 1987), and it appears that many of the puzzling issues related to the confinement of ESE clouds pertain also to the scattering region associated with J1819$+$3845.  Specifically, the problem is that regions that are overpressured by at least three orders of magnitude with respect to the ambient ISM seem to persist throughout the ISM and appear highly prevalent: the ESE event rate is 0.013 source$^{-1}$\,yr$^{-1}$ (Fiedler et al.\,1987).  Given this puzzle, it is natural to consider whether we observe evidence for changes in the internal structure of the cloud with time.  


We place limits on the stability of the turbulence within the screen by investigating whether the small-scale, $10^{7-8}\,$m turbulence that is probed by the scintillations evolves significantly with time.  Several arguments suggest that there is little evolution of the structure on these scales during the time in which rapid scintillations were observed, on timescales between tens of minutes and years:
\begin{itemize}
\item There is a strong degree of cross-correlation between the polarized and unpolarized lightcurves, indicating that the scintillation pattern retains its coherence on the timescale required to traverse the angular distance in the source separating these two components, which is of order an hour.
\item There is a similarly strong degree of cross-correlation between the structure at different frequencies which, like the polarized and unpolarized emission, are also separated in sky position.
\item The annual cycle of the variability timescale is observed to persist over many years and, once the scintillations are scaled to a common timescale that removes this annual cycle, the overall character of the scintillations is consistent with no measurable change over the course of at least two years (see Fig.\,12 of Dennett-Thorpe \& de Bruyn 2003). This is also the case for the seven month interval in 2006, some four years later.
\end{itemize}

Given the extreme pressure inferred for this turbulent region on the basis of its high density, we are led to consider what might confine it; i.e. why doesn't the turbulent cloud simply blow up?  One possibility is that the turbulence is confined by a magnetic field, or is at least in pressure balance with it.  This is reasonable to expect, given that the scattering is observed to be highly anisotropic and thus, that the magnetic field may be dynamically important.  

Based on the density derived in eq.\,(\ref{DensityEstimate}), the magnetic field must have a value of 
\begin{eqnarray}
B_{\rm eq} = 58 \, l_0^{1/6} \, T_4^{1/2} \epsilon^{-1/2} \ \left( \frac{\Delta L}{100\,{\rm AU}} \right)^{-1/4} \,\, \mu{\rm G},
\end{eqnarray}
if it is in pressure balance with the plasma.  This, in turn, implies a Faraday depth of
\begin{eqnarray}
{\rm RM} = 2.2 \, l_0^{1/2} \, T_4^{1/2} \epsilon^{-3/2} \left( \frac{\Delta L}{100\,{\rm AU}} \right)^{1/4} \,\, {\rm rad\,m}^{-2}.
\end{eqnarray}
We are thus led to conclude that, if the magnetic field is roughly in pressure balance with the plasma, the scattering material would cause a small but plausibly detectable increment in the rotation measure of any polarized emission that lies behind it.  

\subsection{RM from RM variations and diffuse Galactic polarization}

Given that the screen that covered J1819$+$3845 may be confined by a high magnetic field, we have searched for a change in the rotation measure of J1819$+$3845 between the scintillation and post-scintillation periods. The 6\,cm band is too narrow to set any interesting RM limits so we have confined our analysis to the 21\,cm polarization results.  However, before discussing these results we review the Galactic RM properties in the general direction of J1819$+$3845. 

\begin{figure}
\includegraphics[width=85mm]{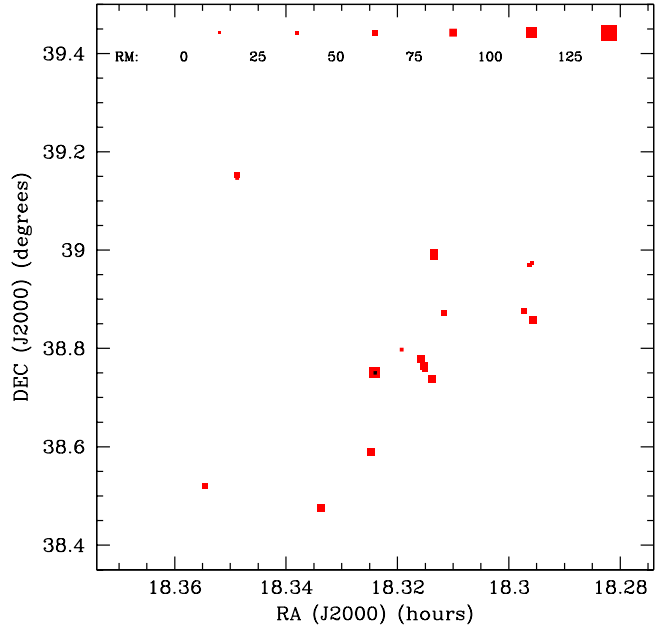}
\caption{A map of the rotation measure field surrounding J1819$+$3845
based on WSRT-selected polarized sources. J1819$+$3845 is marked by
a small black dot. The RM code is shown in the top of the figure.  Many
polarized sources in this image are extended or double, and multiple closely-spaced RM values 
are slightly overlapping.} \label{fig:WSRT_RM}
\end{figure}

We have inspected the RMs of NVSS-selected sources whose polarized flux is larger than 2\,mJy in an area of $8^\circ \times 8^\circ$ centred on J1819$+$3845, taken from Taylor et al.\,(2009).  Most RM values are found between 0 and +50 rad\,m$^{-2}$.  Although there is a conspicuous cloud of higher RM values about 4$^\circ$ south of J1819$+$3845, we have evidence that links this region to the screen towards the quasar. We used the many 12 hour WSRT observations at 21\,cm, in which we can reliably detect sources with polarized fluxes down to 0.1\,mJy, to map the distribution of RMs close to J1819$+$3845. The RMs of those sources are shown in Fig.\,\ref{fig:WSRT_RM} and agree quite well with the range displayed in the NVSS data from Taylor et al.\,(2009). We estimate a value of 50$\pm$10 rad\,m$^{-2}$ for the direction of J1819$+$3845.  There is only one known pulsar  in the direction of J1819$+$3845.   PSR J1813$+$4013 is located 2$^\circ$ away, and has a RM of +47 rad\,m$^{-2}$, in line with many of the background sources.

In total we have more than a dozen good quality syntheses at 21\,cm. In about six of them we have a clear detection of polarized emission from J1819$+$3845 at a typical  level of about 0.5--1\% of the Stokes $I$ flux density.  One of the more interesting cases is shown in Fig.\,\ref{fig:FaradaySpectrum}, data from 21 August 2004, when the source had an unusually high and stable flux density of 240\,mJy.  The average polarized flux density at that epoch was about 1.5\,mJy, peaking at an  RM of +115 rad\,m$^{-2}$.  There is some evidence of variability in the RM of the emission from J1819$+$3845 in the years 2002--2005. If those RM variations are real, they could be due to intrinsic variations in the RM from the quasar, or they may be due to variations in the RM contributed by the screen. For the moment we set a conservative limit of 40 rad\,m$^{-2}$ to the variable component of the RM from J1819$+$3845.    Unfortunately, the two epochs of 21\,cm observations taken in May 2009 and June 2012, when J1819$+$3845 had emerged from behind the screen,  did not reveal any linear polarization in the source, down to limits of about 0.2\%.  (The 21cm $Q$ and $U$ images for the 21 August 2004 epoch also indicate temporal variability during the 12-hour synthesis, with a peak polarized intensity of 3\,mJy.  In view of the 6\,cm polarization results described above, we may also expect significant time delays between the Stokes $Q$ and $U$ signals.  We note in passing that  these time delays will complicate traditional RM synthesis.  We defer a discussion of this topic to a future paper.)

\begin{figure}
\includegraphics[width=75mm,angle=-90]{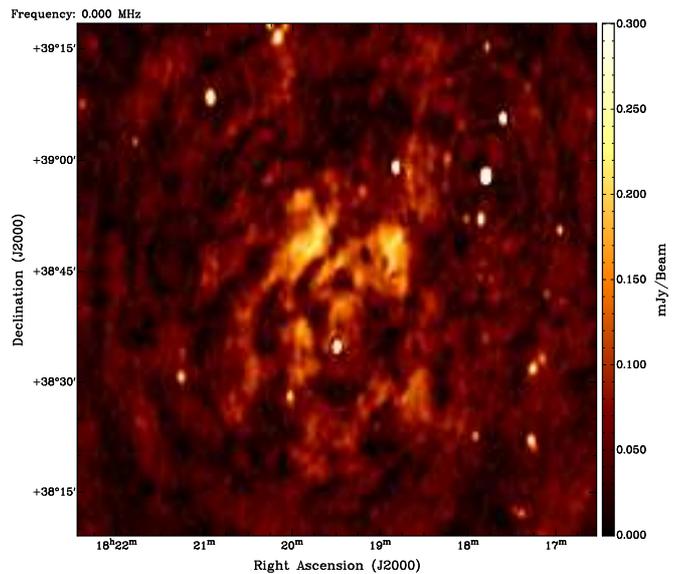}
\caption{Image of the polarized intensity at a Faraday depth of +60 
rad\,m$^{-2}$. J1819$+$3845 is located in the centre of the image.} 
\label{fig:PI60}
\end{figure}

\begin{figure}
\includegraphics[width=75mm,angle=-90]{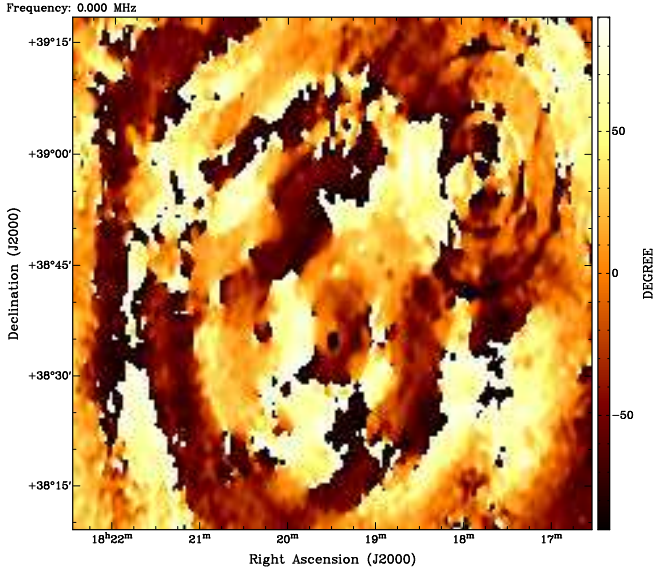}
\caption{Image of the polarization angle at a Faraday depth of +60\,rad\,m$^{-2}$. J1819$+$3845 is located in the centre of the image near the white spot.}  
\label{fig:PA60}
\end{figure}

The WSRT 21cm data have also been used to image the diffuse polarized
emission from the Galactic foreground. We used four 12-hour syntheses with complete short spacing coverage with an 18 m increment  to avoid confusion.  The surface brightness of the
diffuse polarization is, however, very low and we therefore smoothed the data to
a resolution of 50$^{\prime \prime} \times$ 80$^{\prime \prime}$.  RM synthesis was used to make a cube of images running from -500 to +500 rad m$^{-2}$ in increments of 20 rad m$^{-2}$.  The polarized intensity and polarization angle for the frame at Faraday depth +60 rad\,m$^{-2}$ are shown in Figs.\,\ref{fig:PI60} and \ref{fig:PA60}, where the bulk of the emission peaks.  Although the RMSF has a width of 350 rad m$^{-2}$, the S/N is sufficient to establish that there are detectable spatial variations of the peak Faraday depth between 40 and 80 rad\,m$^{-2}$. The diffuse polarized emission is confined to the innermost area defined by the primary beam of the WSRT 25\,m dishes (the HPBW at 21\,cm is 0.6$^\circ$).   
Most of the discrete sources visible in the inner 30$^{\prime}$ of the image are intrinsically polarized. Their RM values are included in Fig.\,\ref{fig:WSRT_RM}.   The morphology of the polarized emission is quite complicated, and there is no clear spatial morphological connection to the quasar. This, of course, is not surprising because the diffuse foreground polarization is built up over tens, if not hundreds, of parsecs in the Galactic disk and halo and must therefore originate predominately behind the screen.  

The polarization angle of the diffuse emission surrounding J1819$+$3845 is smoothly distributed. No distinct small-scale structures can be seen that could be ascribed to the scintillation screen which, at the time of these observations, would still include the position of the quasar.  Because we know that one edge of the screen must currently (as of September 2012) be less then 10--20$\arcsec$ from the location of the quasar, we can set a conservative  upper limit of about 0.5 radians to the change in polarization of the background diffuse emission induced by the screen. This translates to a RM limit of about 10 rad\,m$^{-2}$. 

We also used the two 12\,h  syntheses with the WSRT in the 315--385\,MHz band taken in 2004 and 2005  to search for diffuse  polarized emission. There is definitely diffuse polarization at a Faraday depth of about 40--50 rad\,m$^{-2}$ in an area of 2$^\circ$ diameter surrounding J1819$+$3845.
That is, the polarized emission seen at 21\,cm probably extends well beyond the field of view that can be imaged at 21\,cm. Unfortunately, the short spacing uv-coverage in the 350\,MHz observation is insufficient to reliably image the full field of view; i.e. the emission is self-confused.   We plan new observations at 315--385\,MHz with better uv-coverage to image this emission. This could provide an order of magnitude improvement in sensitivity to RM structure in the area around J1819$+$3845 compared to the limits set by the 21\,cm observations described above. 



\section{Conclusion} \label{sec:con}

We have presented a comprehensive summary of the multi-frequency monitoring of the dramatic scintillating radio source J1819$+$3845, over a period of 13 years.  A unique dataset, unparalleled in the amount of detail, 
cadence of monitoring and accuracy in both total intensity and polarization lightcurves has allowed us to make
far-reaching   deductions about the properties of the source and screen.  The major results and conclusions are as follows.

\begin{itemize}
\item For more than 7.5 years the quasar J1819$+$3845 was the most dramatically scintillating extragalactic radio source in the sky. 
\item Sometime between June 2006 and February 2007, the scintillations in the quasar J1819$+$3845 vanished, and the source now exhibits variations at a level below a modulation index of $m=1$\%.
\item Neither the source flux density nor spectrum indicate abnormal behaviour before or after the disappearance of the scintillations.
\item The VLBI structure of the quasar remains compact in the post-scintillation period, and this unequivocally demonstrates that the disappearance of the scintillations can not be due to the source, but rather must be attributed to the passage of the nearby (1--2\,pc distant) scattering screen out of the line of sight to the quasar.
\item We have presented, for the first time, detailed polarization lightcurves of the source.  The variations in $Q$ and $U$ are generally well-correlated with the fluctuations in $I$, although the nature of the correlation does change over time. 
Depending on the effective scintillation speed, $Q$ and $U$ lag $I$ by typically an hour, indicating that there is an angular offset in the centroid of the polarized emission with respect to the total intensity.  We have analysed these offsets over multiple epochs to derive a typical speed at which polarized ejecta propagate along the jet internal to the source of $\sim 0.3\,c$. 
\item The disappearance of the scintillations in J1819$+$3845 enables us to find that the edge of the scattering cloud has a width less than 6\,AU and a transverse extent of at least 59\,AU.  Given the known scintillation velocity of the scattering cloud and the date at which scintillations ceased, the location of the cloud edge can be calculated for any given epoch.  This location is useful for definitively identifying signatures of the cloud in future observations.
\item The large scattering measure of the turbulent cloud formerly responsible for the scattering towards J1819$+$3845 must have a high density.  We derive a density of $n_e = 97 \, l_0^{1/3} \epsilon^{-1}  \left( \Delta L/100\,{\rm AU} \right)^{-1/2} \, {\rm cm}^{-3}$.  The associated DM increment is $4.7 l_0^{1/3} \epsilon^{-1} \left( \Delta L/100\,{\rm AU} \right)^{1/2}\,$pc\,cm$^{-3}$.
\item If the magnetic pressure is comparable to the thermal pressure inside the turbulent region, one expects a magnetic field of $\sim 58\,\mu$G internal to the scattering region.  Such a magnetic field, combined with the high local plasma density, should produce a $\sim 2\,$rad\,m$^{-2}$ increment in the rotation measure.  
\item We present images of the 21\,cm-wavelength diffuse polarization in the region surrounding J1819$+$3845.  We place a conservative lower limit of 10\,rad\,m$^{-2}$ on any possible RM change associated with the region near J1819$+$3845.  Because the diffuse emission observed at 21\,cm is built up over many tens to hundreds of parsecs in the ISM, it is not surprising that no obvious evidence of the scattering region is observed.  We expect future observations at lower frequencies to either detect or place much more stringent constraints on the RM of the scattering region.
\item We present a map of the RMs of polarized background sources in the vicinity of J1819$+$3845.  There is a region of large RMs $\sim 4^\circ$ to the South of the quasar, but the density of sources across the region is too low at present to draw firm conclusions on the nature of the RM field in the direct vicinity of the scattering screen.
\item We have examined other compact sources in the vicinity of J1819$+$3845 for similarly fast variations that may indicate that they are being affected by the same scattering material as J1819$+$3845.  No other source in the 21\,cm field exhibits intra-day variability.  Two sources exhibit relatively high modulations of 11\% and 12\%, but their slow characteristic timescale suggests that they are exhibiting ISS from a far more distant scattering screen.
\item It is striking to compare the anomalous properties of the scattering screen formerly responsible for the variations in J1819$+$3845, particularly its high electron density, with other anomalous scattering features in the ISM, most notably ESEs (Fiedler et al. 1987; Romani et al. 1987). We note that the one existing limit on the RM change associated with an ESE in front of the quasar PKS\,1741--038 of $|\Delta {\rm RM}| \leq 10.1\,$rad\,m$^{-2}$ (Clegg et al.\,1996) places a limit on the ambient magnetic field of $\langle B_\parallel \rangle < 12\,$mG (Lazio, Fey \& Gaume 2000), where the estimated column depth is $N_0 = 10^{-4}\,$pc\,cm$^{-3}$.   The corresponding magnetic field limit for the scattering material towards J1819$+$3845 is derived by taking the estimated electron density based on the scattering measure increment (viz. eq.\,(\ref{DensityEstimate})), together with the limit of $\Delta {\rm RM} < 10\,$rad\,m$^{-2}$, and yields $\langle B_\parallel \rangle < 0.26 \, \epsilon\, l_0^{-1/3} \Delta L_{100}^{-1/2}\,$mG.   However, this limit is still an order of magnitude larger than the field strength required for the thermal pressure to maintain balance with the magnetic field in the case of scattering material toward J1819$+$3845.
\item Finally, we note that the screen in the direction of J1819$+$3845 may be  the nearest object to the Sun and Earth, next to Proxima Centauri.  
\end{itemize}

\begin{acknowledgements}  
The WSRT is operated by the Netherlands Foundation for Research in Astronomy (NFRA/ASTRON)  with financial support by the Netherlands Organization for Scientific Research (NWO).  We thank Dr. Denise Gabuzda for a copy of the thesis of Dr. Brian Moloney.  Parts of this research were conducted by the Australian Research Council Centre of Excellence for All-sky Astrophysics (CAASTRO), through project number CE110001020.
\end{acknowledgements}

\end{document}